\documentclass[
superscriptaddress,
amsmath,amssymb,
aps,
prl,
twocolumn
]{revtex4-1}

\usepackage{hyperref}
\usepackage{graphicx}
\usepackage{color}
\graphicspath{{figures/}}

\newcommand{\sectionprl}[1]{\medskip{\em #1}\/.\,---\,}

\newcommand{\rme}{{\rm e}}
\newcommand{\rmd}{{\rm d}}
\newcommand{\rmi}{{\rm i}}

\begin{document}
\newcommand{\titlename}{Dissipative quantum dynamics, phase transitions and non-Hermitian random matrices}
\title{\titlename}

\author{Mahaveer Prasad}
\email{mahaveer.prasad@icts.res.in}
 \affiliation{International Centre for Theoretical Sciences, Tata Institute of 
Fundamental Research, 560089 Bangalore, India}
\author{Hari Kumar Yadalam}
 \email{hari.kumar@icts.res.in}
\affiliation{International Centre for Theoretical Sciences, Tata Institute of 
Fundamental Research, 560089 Bangalore, India}
\affiliation{Laboratoire de Physique, \'Ecole Normale Sup\'erieure, CNRS, Universit\'e
PSL, Sorbonne Universit\'e, Universit\'e de Paris, 75005 Paris, France}
\author{Camille Aron}
 \email{aron@ens.fr}
\affiliation{Laboratoire de Physique, \'Ecole Normale Sup\'erieure, CNRS, Universit\'e
PSL, Sorbonne Universit\'e, Universit\'e de Paris, 75005 Paris, France}
\author{Manas Kulkarni}
 \email{manas.kulkarni@icts.res.in}
\affiliation{International Centre for Theoretical Sciences, Tata Institute of 
Fundamental Research, 560089 Bangalore, India}

\date{\today}
	
\begin{abstract}
We explore the connections between dissipative quantum phase transitions and non-Hermitian random matrix theory. For this, we work in the framework of the dissipative Dicke model
which is archetypal of symmetry-breaking phase transitions in open quantum systems.
We establish that the Liouvillian describing the quantum dynamics exhibits distinct spectral features of integrable and chaotic character on the two sides of the critical point. 
We follow the distribution of the spacings of the complex Liouvillian eigenvalues across the critical point. 
In the normal and superradiant phases, the distributions are $2D$ Poisson and that of the Ginibre Unitary random matrix ensemble, respectively. 
Our results are corroborated by computing a recently introduced complex-plane generalization of the consecutive level-spacing ratio distribution. 
Our approach can be readily adapted for classifying the nature of quantum dynamics across dissipative critical points in other open quantum systems. 
\end{abstract}


\maketitle

\sectionprl{Introduction}
The notions of integrability and chaos are well formulated for classical interacting (non-linear) systems~\cite{reichl1992transition,Strogatz1996chaos,goldstein2002classical,arnold2013mathematical,lakshmanan2012nonlinear,thompson2002nonlinear, zaslavsky2004hamiltonian,alligood2000chaos,tel2006chaotic,cvitanovic2005Chaos}. 
Similar concepts for quantum mechanical systems have not reached the same level of maturity. 
Classically, integrable versus chaotic features are typically diagnosed by computing Lyapunov exponents~\cite{lyapunov1992general,eckmann1985ergodic,arnold1986lyapunov,lakshmanan2012nonlinear, tel2006chaotic} or by establishing the existence of an extensive number of independent conserved quantities (Liouville integrability)~\cite{olshanetsky1981classical,das1989integrable,babelon2003introduction,zaslavsky2004hamiltonian,arutyunov2019elements,sardanashvilyintegrable}. 
Attempts to generalise these diagnostics to the quantum realm have led, on the one hand, to the definition of Lyapunov exponents from the exponential growth of out-of-time-order correlators (OTOC)~\cite{larkin1969quasiclassical,maldacena2016bound} and, on the other hand, to identifying sets of commuting operators.
The presence, or lack thereof, of an extensive number of these operators manifests itself in the statistical features of the spectrum of the Hamiltonian~\cite{mehta2004random,akemann2011oxford,reichl1992transition,guhr1998random,haake1991quantum, stockmann1999chaos, gutzwiller2013chaos}. 

The spectra of quantum Hamiltonians were conjectured to typically exhibit two distinct behaviors depending on whether their corresponding classical limits are integrable or chaotic.
Initially, Berry and Tabor~\cite{berry1977level} speculated that typical quantum Hamiltonians with an integrable classical limit (except for a few pathological cases) have consecutive level spacings distributed according to the Poisson distribution.
Later, Bohigas, Giannoni and Schmit~\cite{bohigas1984characterization} further conjectured that those Hamiltonians with a chaotic classical limit have spectra exhibiting strong level repulsion and the consecutive level spacings are distributed according to Hermitian random matrix theory (RMT).
Subsequent works have shown that these conjectures are also applicable to those quantum systems that do not have a well-defined classical limit~\cite{brody1981random, hsu1993level, zelevinsky1996nuclear,  kota2001random, santos2004integrability, kudo2005level, santos2010onset, pal2010many, gubin2012quantum, abanin2019many}.

These ideas were later extended to the case of Markovian open quantum systems. 
Instead of studying the spectrum of the isolated Hamiltonian $H$, \textit{i.e.} the generator of closed quantum dynamics, one may study the spectrum of the Liouvillian $\mathcal{L}$, \textit{i.e.} the generator of the evolution of the density matrix $\rho$, $\partial_t \rho = \mathcal{L} \rho $.
Here, $\mathcal{L}$ accounts for both the unitary evolution and for the driven-dissipative processes induced by the coupling to the environment~\cite{breuer2002theory}.
Generically, Liouvillians are non-Hermitian operators with complex spectra. Grobe, Haake, and Sommers~\cite{grobe1988quantum} conjectured that those Liouvillians whose corresponding classical dynamics are integrable have complex level spacings distributed according to the $2D$ Poisson distribution.
On the contrary, those whose classical limits are chaotic follow the predictions from non-Hermitian RMT, specifically from the Ginibre ensembles~\cite{ginibre1965statistical,mehta2004random,forrester2010log}.
More recently, these conjectures were also found to be valid for systems without a well-defined classical limit~\cite{hamazaki2019non,akemann2019universal}. 

The simple intuition behind those successful conjectures goes as follows. For integrable dynamics with an extensive number of commuting conserved quantities, the spectrum of the Hamiltonian or the Liouvillian is expected to be the direct sum of extensively many independent sectors of the theory.
This independence guarantees that levels within different sectors can overlap.
On the other hand, for chaotic dynamics, only a few conserved quantities exist and level repulsion is expected in each symmetry sector.
In turn, this presence or lack thereof of an extensive number of conserved quantities is expected to be reflected in the nearest-level spacing statistics of the spectra.
 
In this work, we study the spectral properties of the Liouvillian of a dissipative version of the paradigmatic Dicke model~\cite{dicke1954coherence,gross1982superradiance}. 
In the thermodynamic limit, the isolated Dicke model displays a $\mathbb{Z}_{2}$ symmetry-breaking quantum phase transition between a normal and a superradiant phase~\cite{garraway2011dicke,hepp1973superradiant,kirton2019introduction}.
Studies of the spectral statistics of the Dicke Hamiltonian~\cite{emary2003chaos1, emary2003chaos2} revealed that the level spacings are Poisson distributed in the normal phase, reflective of integrable dynamics, whereas they are distributed according to the Gaussian Orthogonal Ensemble in the superradiant phase, indicating chaotic dynamics. 
Here, via exact diagonalization of the Liouvillian, we discuss whether and how these connections with RMT can be generalized to the context of phase transitions in open quantum systems.
More precisely, we address the robustness of the signatures of integrability as the system is driven through a phase transition by turning on an integrability-breaking term.

\sectionprl{Dissipative Dicke model}
The dissipative Dicke model describes the coupling of an ensemble of closely packed quantum emitters to a single leaky cavity mode~\cite{dicke1954coherence,gross1982superradiance}. In the Markovian approximation, the evolution of the density matrix $\rho$ is governed by a Lindblad Master equation where the Liouvillian superoperator reads
\begin{align} \label{eq:Liouvillian}
\mathcal{L}\, \star =-\rmi\left[H,\star\right]+\kappa\left[2 a_{}^{} \star 
a^{\dagger}-\left\{a^{\dagger}a,\star\right\}\right],
\end{align}
with $\kappa > 0$ being the cavity decay rate and $\star$ stands for operators on the Hilbert space. The Dicke Hamiltonian is given by
\begin{align}
H = \omega_{\rm c} a^{\dagger}a + \omega_{\rm s} S^{z}+\frac{2 
\lambda}{\sqrt{S}}(a^{\dagger}+a)S^{x}\,.
\end{align}
$a$ ($a^{\dagger}$) is the bosonic annihilation (creation) operator of a cavity mode with energy $\omega_{\rm c}$. $S^{\alpha}$, $\alpha=x,y,z$, are the spin angular momentum operators built from the totally symmetric representation of $S$ identical two-level systems with energy splitting $\omega_{\rm s}$. $\lambda$ is the cavity-spin coupling which is rescaled by $1/\sqrt{S}$ to ensure a non-trivial thermodynamic limit ($S \to \infty$). 
The Dicke Hamiltonian is $\mathbb{Z}_2$-symmetric: $[ H, \Pi ] = 0$ where the operator $\Pi = \exp\left[\rmi \pi (a^\dagger a + S^z +S/2)\right]$ gives the parity of the total number of excitations.
As a consequence of the specific structure of the dissipator in Eq.~(\ref{eq:Liouvillian}), the Liouvillian inherits a so-called weak $\mathbb{Z}_2$ symmetry: $[ \mathcal{L}, \varPi] = 0$ where $\varPi \, \star = \Pi \, \star \, \Pi^\dagger$ gives the parity of the difference of the number of excitations between the left and right sides of the states in Liouville space~\cite{albert2014symmetries,buvca2012note,lieu2020symmetry, SM}.
In the thermodynamic limit, this weak $\mathbb{Z}_{2}$ symmetry is spontaneously broken in the steady state at $\lambda = \lambda^{*} = \frac{1}{2}\sqrt{\omega_{\rm c} \omega_{\rm s}} \sqrt{1+\kappa^2/\omega_{\rm c}^2}$, corresponding to a second-order dissipative phase transition~\cite{Dimer2007Dicke,kirton2019introduction,roses2020dicke}. In 
the normal phase, \textit{i.e.} $\lambda < \lambda^*$, the boson expectation value vanishes: $\langle a \rangle=0$. In the superradiant phase, \textit{i.e.} $\lambda > \lambda^*$, it acquires a finite expectation value: $\langle a \rangle \neq 0$.
At $\lambda=0$, the model is trivially integrable.

\sectionprl{Complex spectra}
We analyze the statistical properties of the complex eigenvalues $\{E_{i}\}$ of the Liouvillian operator $\mathcal{L}$ by means of extensive numerical computations.
We work in the even parity sector of the Liouville space to avoid possible spurious overlaps of eigenvalues from the different symmetry sectors~\cite{rosenzweig1960repulsion}. Throughout the paper, we consider the strongly dissipative regime, $\omega_{\rm c}=\omega_{\rm s}=\kappa=1$, for which the critical point is located at $\lambda^{*}=1/\sqrt{2} \approx 0.71$.

\begin{figure}[!tbp]
\centering
\includegraphics[scale=0.6]
{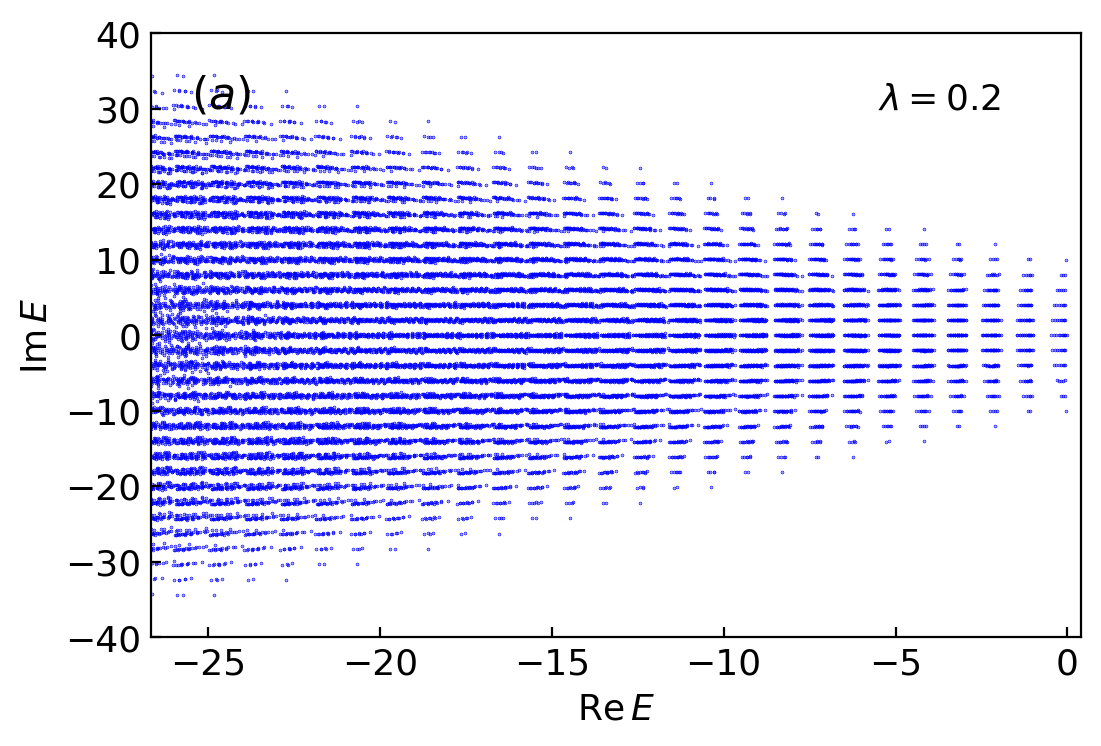}
\includegraphics[scale=0.6]
{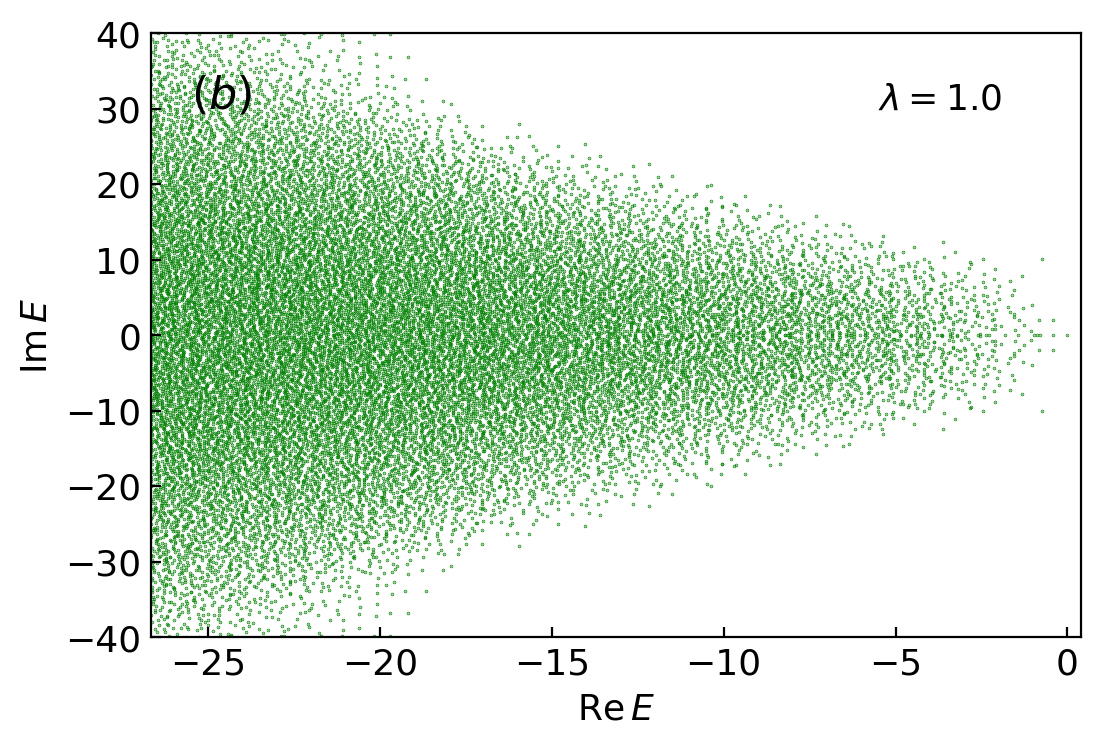}
\caption{Scatter plot of the complex spectrum of the Liouvillian $\mathcal{L}$ of the dissipative Dicke model for $S=10$ and
$\omega_{\rm c}=\omega_{\rm s}=\kappa=1$ for which $\lambda^* = 1/\sqrt{2}$. 
(a) Normal phase, $\lambda=0.2$. 
(b) Superradiant phase, $\lambda=1.0$.
A stark difference in the structure of the spectrum above and below the critical point can be observed.
 }
\label{spectrum_lbd}
\end{figure}

In practice, the numerical approach comes with two inherent 
limitations: 
(\textit{i}) The infinitely large bosonic Hilbert space of the cavity has to be truncated to a finite number of excitations, $n_{\rm cutoff} = 40$.
(\textit{ii}) Numerical errors during the diagonalization process can propagate dangerously, yielding an accuracy of the results far worse than machine precision.
Consequently, we truncate our spectra to an energy window $\mathrm{Re} \, E_{i} \in [- \alpha \kappa\, n_{\rm cuttoff},0]$ where we make sure that statistics are converged with respect to $n_{\rm cutoff}$.
Although its precise value is of little consequence to our findings, we choose $\alpha = 2/3$.
This amounts to analyzing the statistical properties of those eigenvalues which correspond to intermediate to long-lived dynamics.  We work with 128 bits complex double float precision. 

The overall aspect of the spectrum is illustrated in Fig.~\ref{spectrum_lbd} for different values of the cavity-spin coupling $\lambda$ and for fixed spin size $S=10$.
The symmetry about the real axis is a generic feature of Liouvillians of Lindblad Master equations~\cite{haake1991quantum}. The unique steady-state of the dynamics corresponds to the single eigenvalue located at $E = 0$.
The spectra in the two phases display clear differences. 
In the non-interacting limit, $\lambda = 0$, the spectrum displays ladder structures across both the imaginary and the real axis. The former are a direct consequence of our choice of resonant parameters, $\omega_{\rm c} = \omega_{\rm s}$, whereas the latter stem from the fragmentation of the Liouville space due to the presence of continuous symmetries at $\lambda=0$: The weak $U(1)$ symmetry corresponding to the conservation of superoperators $[a^{\dagger}a,\star] $ and the strong $U(1)$ symmetry corresponding to the conservation of $S^z $. 
In the normal phase, $0<\lambda<\lambda^*$, the spectrum in Fig.~\ref{spectrum_lbd}a still displays structured patterns that are inherited from the the non-interacting limit. The effect of a small but finite interaction can be seen as a renormalization of $\omega_{\rm c}$, $\omega_{\rm s}$ and $\kappa$ leading to smearing of the patterned spectrum. There, the existence of patterns across the real axis is robust and we suspect them to be rooted in the fragmentation of Liouville space due to the emergence of approximately conserved quantities. This fragmentation disappears as $\lambda$ approaches $\lambda^*$ and the spectrum does not display such signature of emergent conservation laws in the superradiant phase, $\lambda > \lambda^*$.

\sectionprl{Spacing statistics of complex eigenvalues}
In order to unveil the universal features of these complex spectra, we turn to the study of level-spacing statistics.
We first perform an unfolding of the spectrum using standard procedures~\cite{SM}. 
The unfolded spectrum is then used to generate the histogram of the Euclidean distance $s$ between nearest-neighbor eigenvalues in the complex plane, yielding the complex-level spacing distribution $p(s)$.

\begin{figure}
 \centering
\includegraphics[scale=0.6]
{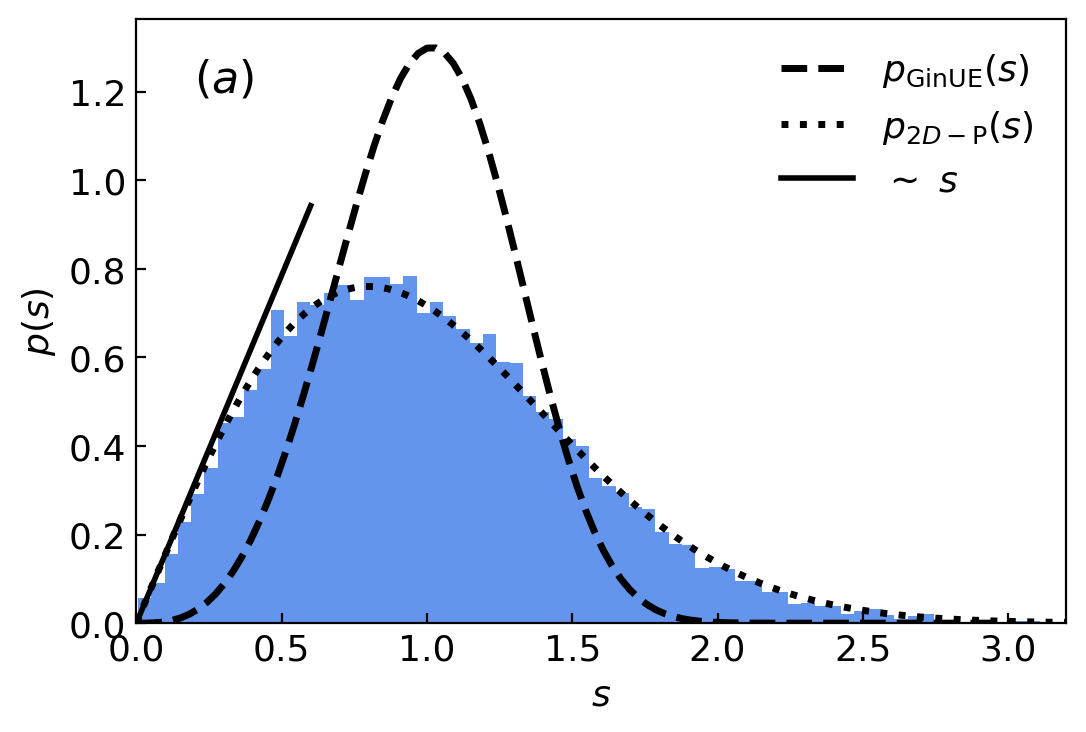}
\includegraphics[scale=0.6]
{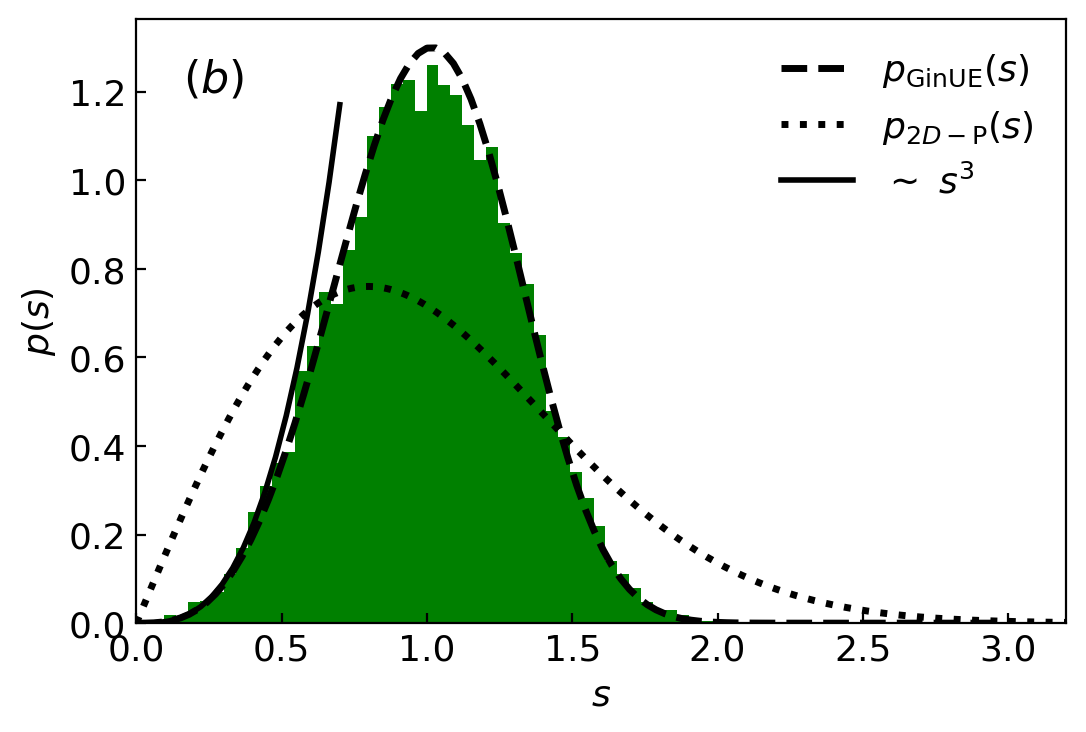}
\caption{Level-spacing distribution of the complex spectrum of the Liouvillian $\mathcal{L}$
in (a) normal phase with $\lambda= 0.2$, and (b) superradiant phase with $\lambda= 1.0$. 
We find remarkable agreement with the $2D$ Poisson distribution $p_{2D\textrm{-P}}(s)$ given in Eq.~(\ref{eq:2dp}) and that of the GinUE RMT prediction $p_{\textrm{GinUE}}(s)$ given in Eq.~(\ref{eq:ginue}) in the normal phase and in the superradiant phase, respectively. }
\label{NSS}
\end{figure}

The results are summarized in Fig.~\ref{NSS} for values of $\lambda$ corresponding to the normal and superradiant phases.
For comparison, we also plot the corresponding spacing distribution for independent complex random numbers, namely the $2D$ Poisson distribution 
\begin{align}
\label{eq:2dp}
p_{2D\textrm{-P}}(s)=\frac{\pi}{2} \, s \, \exp\left( -\pi s^2 /4 \right)\,,
\end{align}
as well as the corresponding distribution for the eigenvalues of non-Hermitian random matrix ensemble~\cite{grobe1988quantum, akemann2019universal}, namely 
the Ginibre Unitary Ensemble (GinUE), 
\begin{align}
\label{eq:ginue}
 p_{\textrm{GinUE}}(s)= \bar{s} \, \bar{p}_{\textrm{GinUE}}(\bar{s} s) \,,
\end{align}
with 
\begin{align}
\bar{p}_{\textrm{GinUE}}(s)= 
\sum \limits_{j=1}^{\infty} \frac{2 s^{2j+1} \exp(-s^2)}{\Gamma(1+j,s^2)}
 \prod \limits_{k=1}^{\infty} \frac{\Gamma(1+k,s^2)}{k!} 
 \, ,
\end{align}
and $\bar{s}= \int_{0}^{\infty} \rmd s \, s \, \bar{p}_{\textrm{GinUE}}(s)$. Here, $\Gamma(1+k,s^2) = \int_{s^2}^{\infty} t^k \rme^{-t } \rmd t$ is the incomplete Gamma function. 
Figure~\ref{NSS} demonstrates that the distributions computed from the spectrum of $\mathcal{L}$ in Eq.~(\ref{eq:Liouvillian}) are in remarkable agreement with $2D$ Poisson in the normal phase, and with the GinUE prediction in the superradiant phase.
In the superradiant phase, this reflects the presence of complex-eigenvalue repulsion characterized by a $p(s) \sim s^3$ suppression at small energy spacings, which is consistent with Eq.~(\ref{eq:ginue}).
On the other hand, in the normal phase we find $p(s) \sim s$, consistent with Eq.~(\ref{eq:2dp}).
This corresponds to the absence of level repulsion in the $2D$ complex plane~\cite{haake1991quantum}. 

\begin{figure}
		\centering
\includegraphics[scale=0.5]
{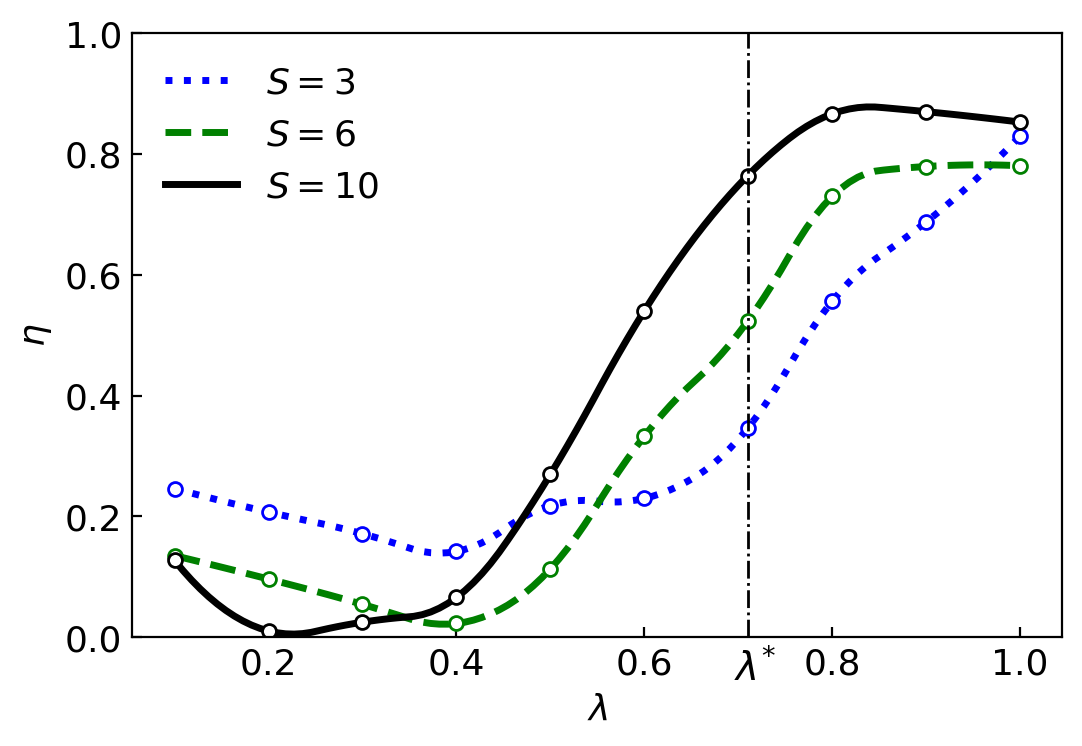}
		\caption{The metric $\eta$ defined in Eq.~(\ref{eq:eta}) as we increase the coupling $\lambda$ from the normal phase to the superradiant phase. It shows the crossover of the complex-eigenvalue spacing distribution from integrable ($\eta \sim 0 $) to RMT ($\eta \sim
1$ ) predictions. The crossover sharpens as we increase the system size. At $\lambda=0$ the dissipative cavity decouples from the spin. Hence, spectrum is expected to display pathological statistics away from any universal behaviour. 
This explains the observed discrepancies close to $\lambda=0$.}
\label{eta_lbd}
\end{figure}

In order to better quantify the nature of the statistics as one crosses from one phase to another, we introduce the metric
\begin{align}
\label{eq:eta}
\displaystyle \eta\equiv \frac{
\int_{0}^{\infty} \rmd s \, [ p(s)-p_{2D\textrm{-P}}(s)]^2}{\int_{0}^{\infty} \rmd s \,[p_{\textrm{GinUE}}(s)-p_{2D\textrm{-P}}(s)]^2}\,.
\end{align}
By construction, $\eta$ vanishes when the numerically obtained distribution $p(s)$ approaches the $2D$ Poisson distribution, whereas $\eta$ goes to $1$ when $p(s)$ approaches the GinUE prediction. Figure~\ref{eta_lbd}, showing $\eta$ versus $\lambda$, exhibits the crossover from a $2D$ Poisson distribution to that of GinUE prediction as one crosses the critical point. This crossover sharpens with increasing the system size.

\begin{figure}[!tbp]
\centering
\vspace{1ex}
\hspace{-2ex}
\includegraphics[scale=0.39]{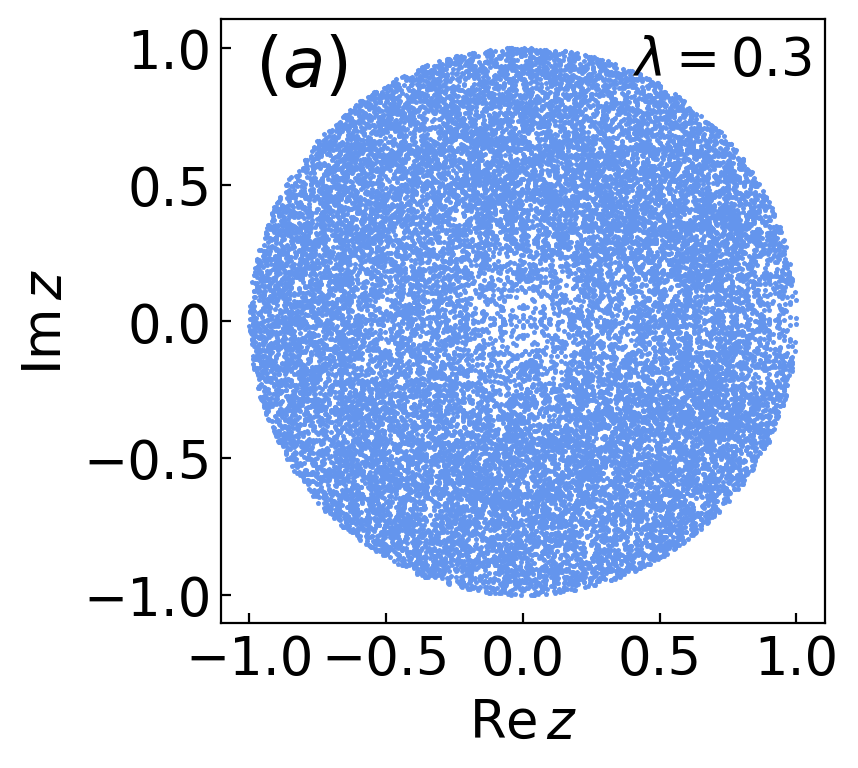}
\hspace{-1ex}
\includegraphics[scale=0.39]{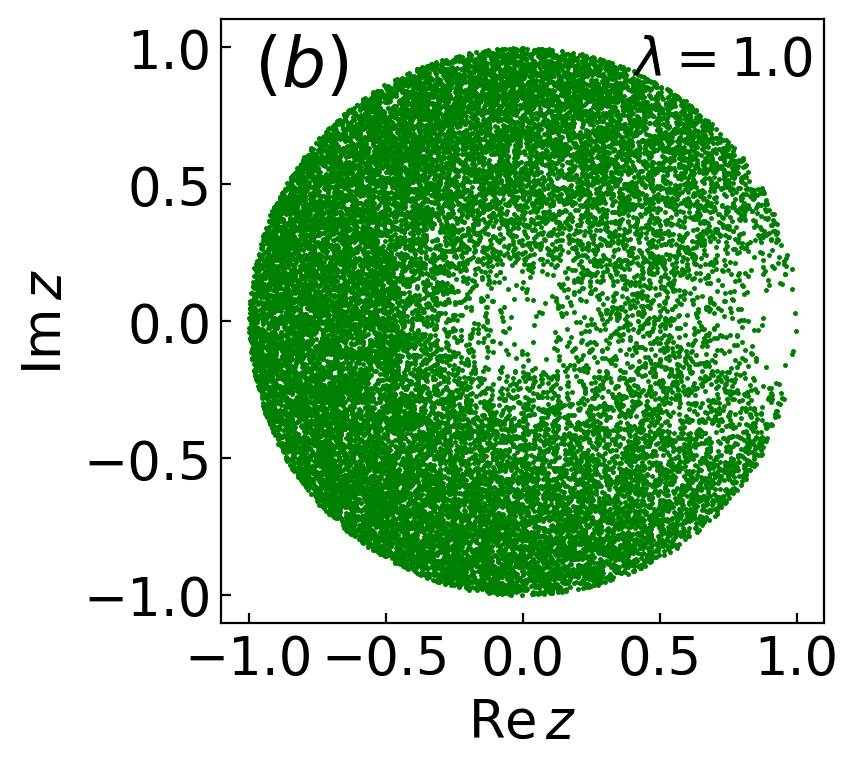}\\
\vspace{1em}
\begin{tabular}{|c|c|c||c|c|}
\hline
 & $\lambda = 0.3 $ & $2D$ Poisson & $\lambda = 1.0 $ &\phantom{aa} GinUE \phantom{aa} \\ 
\hline 
-$\langle \cos \theta \rangle$ & 0.00 & 0 & 0.24 & 0.24 \\ 
\hline 
$\langle r \rangle$ & 0.69 & 0.67 & 0.74 & 0.74 \\
\hline 
\end{tabular} 
\caption{
Scatter plot of the complex level-spacing ratio $z$ introduced in Eq.~(\ref{eq:ratio}) for $S=10$
(a) in the normal phase, $\lambda=0.3$, and
(b) in the superradiant phase, $\lambda=1.0$. 
The table gives the corresponding values of $\langle \cos\theta\rangle$ and $\langle r\rangle$, along with their prediction from the $2D$ Poisson distribution and GinUE RMT.
 }
\label{fig:scatter_plot}
\end{figure}

\sectionprl{Complex-plane generalization of the consecutive level-spacing ratio}
Until now, we only probed spectral statistics using the Euclidean distance $s$ between complex levels. In order to extract the angular information we resort to a recently introduced diagnostic~\cite{sa2020complex} involving the level-spacing ratio 
\begin{align}
\label{eq:ratio}
z_i= r_i \, \rme^{\rmi\theta_i}= \frac{E_i^{\rm NN}- E_i}{E_i^{\rm NNN}-E_i} \,,
\end{align}
where superscripts NN (NNN) stand for nearest (next-nearest) neighbor.
Equation~(\ref{eq:ratio}) is the generalization of the well-known adjacent gap ratio~\cite{oganesyan2007localization, atas2013distribution} defined for isolated quantum systems.
It captures information about next-nearest neighbors which is missed in the conventional diagnostics of level-spacing statistics.
An additional advantage of this quantity is that it does not rely on the unfolding procedure which may sometimes be ambiguous and unreliable.
In Fig.~\ref{fig:scatter_plot}, we show the scatter plots of $z_i$ below and above the critical point. The anisotropy in the superradiant phase is another signature of connection to RMT~\cite{sa2020complex}.
To quantitatively compare with the predictions of $2D$ Poisson and GinUE RMT, we report the extracted values of $\langle r \rangle$ and $\langle \cos \theta \rangle$ in the table below Fig.~\ref{fig:scatter_plot}.

\sectionprl{Conclusion}
We investigated how the presence of a dissipative quantum phase transition driven by an integrability-breaking term affects the spectral statistics of the complex Liouvillian spectrum of open quantum systems. The naive expectation would be to find chaotic features as soon as integrability is broken. Working in the framework of the dissipative Dicke model, we found the spectral features of integrability to survive until the dissipative quantum phase transition ($\lambda < \lambda^*$), after which they are eventually replaced by RMT features indicative of chaotic dynamics ($\lambda > \lambda^*$).
The approach we developed here can be straightforwardly adapted to other dissipative quantum dynamics. This will be instrumental to further establish whether this robustness of integrability features against integrability-breaking terms is a general trait in the context of open quantum systems.

\sectionprl{Acknowledgements}
HKY, CA, MK are grateful for the support from the project 6004-1 of the Indo-French Centre for the Promotion of Advanced Research (IFCPAR). MK acknowledges the support of the Ramanujan Fellowship (SB/S2/RJN-114/2016), SERB Early Career Research Award (ECR/2018/002085) and SERB Matrics Grant (MTR/2019/001101) from the Science and Engineering Research Board (SERB), Department of Science and Technology, Government of India. MK acknowledges support of the Department of Atomic Energy, Government of India, under Project No. RTI4001. CA acknowledges the support from the French ANR ``MoMA'' project ANR-19-CE30-0020. MK thanks the hospitality of  \'Ecole Normale Sup\'erieure (Paris).

\bibliography{references.bib}

\begin{thebibliography}{61}%
\makeatletter
\providecommand \@ifxundefined [1]{%
 \@ifx{#1\undefined}
}%
\providecommand \@ifnum [1]{%
 \ifnum #1\expandafter \@firstoftwo
 \else \expandafter \@secondoftwo
 \fi
}%
\providecommand \@ifx [1]{%
 \ifx #1\expandafter \@firstoftwo
 \else \expandafter \@secondoftwo
 \fi
}%
\providecommand \natexlab [1]{#1}%
\providecommand \enquote  [1]{``#1''}%
\providecommand \bibnamefont  [1]{#1}%
\providecommand \bibfnamefont [1]{#1}%
\providecommand \citenamefont [1]{#1}%
\providecommand \href@noop [0]{\@secondoftwo}%
\providecommand \href [0]{\begingroup \@sanitize@url \@href}%
\providecommand \@href[1]{\@@startlink{#1}\@@href}%
\providecommand \@@href[1]{\endgroup#1\@@endlink}%
\providecommand \@sanitize@url [0]{\catcode `\\12\catcode `\$12\catcode
  `\&12\catcode `\#12\catcode `\^12\catcode `\_12\catcode `\%12\relax}%
\providecommand \@@startlink[1]{}%
\providecommand \@@endlink[0]{}%
\providecommand \url  [0]{\begingroup\@sanitize@url \@url }%
\providecommand \@url [1]{\endgroup\@href {#1}{\urlprefix }}%
\providecommand \urlprefix  [0]{URL }%
\providecommand \Eprint [0]{\href }%
\providecommand \doibase [0]{http://dx.doi.org/}%
\providecommand \selectlanguage [0]{\@gobble}%
\providecommand \bibinfo  [0]{\@secondoftwo}%
\providecommand \bibfield  [0]{\@secondoftwo}%
\providecommand \translation [1]{[#1]}%
\providecommand \BibitemOpen [0]{}%
\providecommand \bibitemStop [0]{}%
\providecommand \bibitemNoStop [0]{.\EOS\space}%
\providecommand \EOS [0]{\spacefactor3000\relax}%
\providecommand \BibitemShut  [1]{\csname bibitem#1\endcsname}%
\let\auto@bib@innerbib\@empty
\bibitem [{\citenamefont {Reichl}(1992)}]{reichl1992transition}%
  \BibitemOpen
  \bibfield  {author} {\bibinfo {author} {\bibfnamefont {L.~E.}\ \bibnamefont
  {Reichl}},\ }\href@noop {} {\emph {\bibinfo {title} {The transition to
  chaos}}},\ Vol.~\bibinfo {volume} {6}\ (\bibinfo  {publisher} {Springer},\
  \bibinfo {year} {1992})\BibitemShut {NoStop}%
\bibitem [{\citenamefont {Strogatz}(1996)}]{Strogatz1996chaos}%
  \BibitemOpen
  \bibfield  {author} {\bibinfo {author} {\bibfnamefont {S.~H.}\ \bibnamefont
  {Strogatz}},\ }\href@noop {} {\emph {\bibinfo {title} {Nonlinear Dynamics and
  Chaos}}}\ (\bibinfo  {publisher} {Perseus Books},\ \bibinfo {year}
  {1996})\BibitemShut {NoStop}%
\bibitem [{\citenamefont {Goldstein}\ \emph {et~al.}(2002)\citenamefont
  {Goldstein}, \citenamefont {Poole},\ and\ \citenamefont
  {Safko}}]{goldstein2002classical}%
  \BibitemOpen
  \bibfield  {author} {\bibinfo {author} {\bibfnamefont {H.}~\bibnamefont
  {Goldstein}}, \bibinfo {author} {\bibfnamefont {C.}~\bibnamefont {Poole}}, \
  and\ \bibinfo {author} {\bibfnamefont {J.}~\bibnamefont {Safko}},\
  }\href@noop {} {\emph {\bibinfo {title} {Classical Mechanics}}}\ (\bibinfo
  {publisher} {Addison Wesley},\ \bibinfo {year} {2002})\BibitemShut {NoStop}%
\bibitem [{\citenamefont {Arnol'd}(2013)}]{arnold2013mathematical}%
  \BibitemOpen
  \bibfield  {author} {\bibinfo {author} {\bibfnamefont {V.~I.}\ \bibnamefont
  {Arnol'd}},\ }\href@noop {} {\emph {\bibinfo {title} {Mathematical methods of
  classical mechanics}}},\ Vol.~\bibinfo {volume} {60}\ (\bibinfo  {publisher}
  {Springer Science \& Business Media},\ \bibinfo {year} {2013})\BibitemShut
  {NoStop}%
\bibitem [{\citenamefont {Lakshmanan}\ and\ \citenamefont
  {Rajaseekar}(2012)}]{lakshmanan2012nonlinear}%
  \BibitemOpen
  \bibfield  {author} {\bibinfo {author} {\bibfnamefont {M.}~\bibnamefont
  {Lakshmanan}}\ and\ \bibinfo {author} {\bibfnamefont {S.}~\bibnamefont
  {Rajaseekar}},\ }\href@noop {} {\emph {\bibinfo {title} {Nonlinear dynamics:
  integrability, chaos and patterns}}}\ (\bibinfo  {publisher} {Springer
  Science \& Business Media},\ \bibinfo {year} {2012})\BibitemShut {NoStop}%
\bibitem [{\citenamefont {Thompson}\ and\ \citenamefont
  {Stewart}(2002)}]{thompson2002nonlinear}%
  \BibitemOpen
  \bibfield  {author} {\bibinfo {author} {\bibfnamefont {J.}~\bibnamefont
  {Thompson}}\ and\ \bibinfo {author} {\bibfnamefont {H.}~\bibnamefont
  {Stewart}},\ }\href@noop {} {\emph {\bibinfo {title} {Nonlinear Dynamics and
  Chaos}}}\ (\bibinfo  {publisher} {Wiley},\ \bibinfo {year}
  {2002})\BibitemShut {NoStop}%
\bibitem [{\citenamefont {Zaslavsky}(2004)}]{zaslavsky2004hamiltonian}%
  \BibitemOpen
  \bibfield  {author} {\bibinfo {author} {\bibfnamefont {G.~M.}\ \bibnamefont
  {Zaslavsky}},\ }\href@noop {} {\emph {\bibinfo {title} {Hamiltonian chaos and
  fractional dynamics}}}\ (\bibinfo  {publisher} {Oxfoed University Press},\
  \bibinfo {year} {2004})\BibitemShut {NoStop}%
\bibitem [{\citenamefont {Alligood}\ \emph {et~al.}(2000)\citenamefont
  {Alligood}, \citenamefont {Sauer},\ and\ \citenamefont
  {Yorke}}]{alligood2000chaos}%
  \BibitemOpen
  \bibfield  {author} {\bibinfo {author} {\bibfnamefont {K.}~\bibnamefont
  {Alligood}}, \bibinfo {author} {\bibfnamefont {T.}~\bibnamefont {Sauer}}, \
  and\ \bibinfo {author} {\bibfnamefont {J.}~\bibnamefont {Yorke}},\
  }\href@noop {} {\emph {\bibinfo {title} {Chaos: An Introduction to Dynamical
  Systems}}},\ Textbooks in Mathematical Sciences\ (\bibinfo  {publisher}
  {Springer},\ \bibinfo {year} {2000})\BibitemShut {NoStop}%
\bibitem [{\citenamefont {T{\'e}l}\ \emph {et~al.}(2006)\citenamefont
  {T{\'e}l}, \citenamefont {Gruiz}, \citenamefont {Kulacsy},\ and\
  \citenamefont {Hadobas}}]{tel2006chaotic}%
  \BibitemOpen
  \bibfield  {author} {\bibinfo {author} {\bibfnamefont {T.}~\bibnamefont
  {T{\'e}l}}, \bibinfo {author} {\bibfnamefont {M.}~\bibnamefont {Gruiz}},
  \bibinfo {author} {\bibfnamefont {K.}~\bibnamefont {Kulacsy}}, \ and\
  \bibinfo {author} {\bibfnamefont {S.}~\bibnamefont {Hadobas}},\ }\href@noop
  {} {\emph {\bibinfo {title} {Chaotic Dynamics: An Introduction Based on
  Classical Mechanics}}},\ Chaotic Dynamics: An Introduction Based on Classical
  Mechanics\ (\bibinfo  {publisher} {Cambridge University Press},\ \bibinfo
  {year} {2006})\BibitemShut {NoStop}%
\bibitem [{\citenamefont {Cvitanovic}\ \emph {et~al.}()\citenamefont
  {Cvitanovic}, \citenamefont {Artuso}, \citenamefont {Mainieri}, \citenamefont
  {Tanner},\ and\ \citenamefont {Vattay}}]{cvitanovic2005Chaos}%
  \BibitemOpen
  \bibfield  {author} {\bibinfo {author} {\bibfnamefont {P.}~\bibnamefont
  {Cvitanovic}}, \bibinfo {author} {\bibfnamefont {R.}~\bibnamefont {Artuso}},
  \bibinfo {author} {\bibfnamefont {R.}~\bibnamefont {Mainieri}}, \bibinfo
  {author} {\bibfnamefont {G.}~\bibnamefont {Tanner}}, \ and\ \bibinfo {author}
  {\bibfnamefont {G.}~\bibnamefont {Vattay}},\ }\href@noop {} {\emph {\bibinfo
  {title} {Chaos: Classical and Quantum}}}\BibitemShut {NoStop}%
\bibitem [{\citenamefont {Lyapunov}(1992)}]{lyapunov1992general}%
  \BibitemOpen
  \bibfield  {author} {\bibinfo {author} {\bibfnamefont {A.~M.}\ \bibnamefont
  {Lyapunov}},\ }\href {\doibase 10.1080/00207179208934253} {\bibfield
  {journal} {\bibinfo  {journal} {Int. J. Control}\ }\textbf {\bibinfo {volume}
  {55}},\ \bibinfo {pages} {531} (\bibinfo {year} {1992})}\BibitemShut
  {NoStop}%
\bibitem [{\citenamefont {Eckmann}\ and\ \citenamefont
  {Ruelle}(1985)}]{eckmann1985ergodic}%
  \BibitemOpen
  \bibfield  {author} {\bibinfo {author} {\bibfnamefont {J.~P.}\ \bibnamefont
  {Eckmann}}\ and\ \bibinfo {author} {\bibfnamefont {D.}~\bibnamefont
  {Ruelle}},\ }\href {\doibase 10.1103/RevModPhys.57.617} {\bibfield  {journal}
  {\bibinfo  {journal} {Rev. Mod. Phys.}\ }\textbf {\bibinfo {volume} {57}},\
  \bibinfo {pages} {617} (\bibinfo {year} {1985})}\BibitemShut {NoStop}%
\bibitem [{\citenamefont {Arnold}\ and\ \citenamefont
  {Wihstutz}()}]{arnold1986lyapunov}%
  \BibitemOpen
  \bibfield  {author} {\bibinfo {author} {\bibfnamefont {L.}~\bibnamefont
  {Arnold}}\ and\ \bibinfo {author} {\bibfnamefont {V.}~\bibnamefont
  {Wihstutz}},\ }\href@noop {} {\emph {\bibinfo {title} {Lyapunov
  Exponents}}},\ Lecture Notes in Mathematics\BibitemShut {NoStop}%
\bibitem [{\citenamefont {Olshanetsky}\ and\ \citenamefont
  {Perelomov}()}]{olshanetsky1981classical}%
  \BibitemOpen
  \bibfield  {author} {\bibinfo {author} {\bibfnamefont {M.~A.}\ \bibnamefont
  {Olshanetsky}}\ and\ \bibinfo {author} {\bibfnamefont {A.~M.}\ \bibnamefont
  {Perelomov}},\ }\href@noop {} {\emph {\bibinfo {title} {Classical integrable
  finite-dimensional systems related to Lie algebras}}}\BibitemShut {NoStop}%
\bibitem [{\citenamefont {Das}(1989)}]{das1989integrable}%
  \BibitemOpen
  \bibfield  {author} {\bibinfo {author} {\bibfnamefont {A.}~\bibnamefont
  {Das}},\ }\href@noop {} {\emph {\bibinfo {title} {Integrable models}}},\
  Vol.~\bibinfo {volume} {30}\ (\bibinfo  {publisher} {World scientific},\
  \bibinfo {year} {1989})\BibitemShut {NoStop}%
\bibitem [{\citenamefont {Babelon}\ \emph {et~al.}(2003)\citenamefont
  {Babelon}, \citenamefont {Bernard},\ and\ \citenamefont
  {Talon}}]{babelon2003introduction}%
  \BibitemOpen
  \bibfield  {author} {\bibinfo {author} {\bibfnamefont {O.}~\bibnamefont
  {Babelon}}, \bibinfo {author} {\bibfnamefont {D.}~\bibnamefont {Bernard}}, \
  and\ \bibinfo {author} {\bibfnamefont {M.}~\bibnamefont {Talon}},\
  }\href@noop {} {\emph {\bibinfo {title} {Introduction to classical integrable
  systems}}}\ (\bibinfo  {publisher} {Cambridge University Press},\ \bibinfo
  {year} {2003})\BibitemShut {NoStop}%
\bibitem [{\citenamefont {Arutyunov}(2019)}]{arutyunov2019elements}%
  \BibitemOpen
  \bibfield  {author} {\bibinfo {author} {\bibfnamefont {G.}~\bibnamefont
  {Arutyunov}},\ }\href@noop {} {\emph {\bibinfo {title} {Elements of classical
  and quantum integrable systems}}}\ (\bibinfo  {publisher} {Springer},\
  \bibinfo {year} {2019})\BibitemShut {NoStop}%
\bibitem [{\citenamefont {Sardanashvily}(2015)}]{sardanashvilyintegrable}%
  \BibitemOpen
  \bibfield  {author} {\bibinfo {author} {\bibfnamefont {G.}~\bibnamefont
  {Sardanashvily}},\ }\href@noop {} {\emph {\bibinfo {title} {Handbook of
  integrable Hamiltonian systems}}}\ (\bibinfo  {publisher} {URRS Moscow},\
  \bibinfo {year} {2015})\BibitemShut {NoStop}%
\bibitem [{\citenamefont {Larkin}\ and\ \citenamefont
  {Ovchinnikov}(1969)}]{larkin1969quasiclassical}%
  \BibitemOpen
  \bibfield  {author} {\bibinfo {author} {\bibfnamefont {A.}~\bibnamefont
  {Larkin}}\ and\ \bibinfo {author} {\bibfnamefont {Y.~N.}\ \bibnamefont
  {Ovchinnikov}},\ }\href
  {http://www.jetp.ras.ru/cgi-bin/e/index/e/28/6/p1200?a=list} {\bibfield
  {journal} {\bibinfo  {journal} {Sov. Phys. JETP}\ }\textbf {\bibinfo {volume}
  {28}},\ \bibinfo {pages} {1200} (\bibinfo {year} {1969})}\BibitemShut
  {NoStop}%
\bibitem [{\citenamefont {Maldacena}\ \emph {et~al.}(2016)\citenamefont
  {Maldacena}, \citenamefont {Shenker},\ and\ \citenamefont
  {Stanford}}]{maldacena2016bound}%
  \BibitemOpen
  \bibfield  {author} {\bibinfo {author} {\bibfnamefont {J.}~\bibnamefont
  {Maldacena}}, \bibinfo {author} {\bibfnamefont {S.~H.}\ \bibnamefont
  {Shenker}}, \ and\ \bibinfo {author} {\bibfnamefont {D.}~\bibnamefont
  {Stanford}},\ }\href@noop {} {\bibfield  {journal} {\bibinfo  {journal} {J.
  High Energy Phys.}\ }\textbf {\bibinfo {volume} {2016}},\ \bibinfo {pages}
  {1} (\bibinfo {year} {2016})}\BibitemShut {NoStop}%
\bibitem [{\citenamefont {Mehta}(2004)}]{mehta2004random}%
  \BibitemOpen
  \bibfield  {author} {\bibinfo {author} {\bibfnamefont {M.~L.}\ \bibnamefont
  {Mehta}},\ }\href@noop {} {\emph {\bibinfo {title} {Random matrices}}}\
  (\bibinfo  {publisher} {Elsevier},\ \bibinfo {year} {2004})\BibitemShut
  {NoStop}%
\bibitem [{\citenamefont {Akemann}\ \emph {et~al.}(2011)\citenamefont
  {Akemann}, \citenamefont {Baik},\ and\ \citenamefont
  {Di~Francesco}}]{akemann2011oxford}%
  \BibitemOpen
  \bibfield  {author} {\bibinfo {author} {\bibfnamefont {G.}~\bibnamefont
  {Akemann}}, \bibinfo {author} {\bibfnamefont {J.}~\bibnamefont {Baik}}, \
  and\ \bibinfo {author} {\bibfnamefont {P.}~\bibnamefont {Di~Francesco}},\
  }\href@noop {} {\emph {\bibinfo {title} {The Oxford handbook of random matrix
  theory}}}\ (\bibinfo  {publisher} {Oxford University Press},\ \bibinfo {year}
  {2011})\BibitemShut {NoStop}%
\bibitem [{\citenamefont {Guhr}\ \emph {et~al.}(1998)\citenamefont {Guhr},
  \citenamefont {Müller–Groeling},\ and\ \citenamefont
  {Weidenmüller}}]{guhr1998random}%
  \BibitemOpen
  \bibfield  {author} {\bibinfo {author} {\bibfnamefont {T.}~\bibnamefont
  {Guhr}}, \bibinfo {author} {\bibfnamefont {A.}~\bibnamefont
  {Müller–Groeling}}, \ and\ \bibinfo {author} {\bibfnamefont {H.~A.}\
  \bibnamefont {Weidenmüller}},\ }\href {\doibase
  https://doi.org/10.1016/S0370-1573(97)00088-4} {\bibfield  {journal}
  {\bibinfo  {journal} {Phys. Rep.}\ }\textbf {\bibinfo {volume} {299}},\
  \bibinfo {pages} {189} (\bibinfo {year} {1998})}\BibitemShut {NoStop}%
\bibitem [{\citenamefont {Haake}(1991)}]{haake1991quantum}%
  \BibitemOpen
  \bibfield  {author} {\bibinfo {author} {\bibfnamefont {F.}~\bibnamefont
  {Haake}},\ }in\ \href@noop {} {\emph {\bibinfo {booktitle} {Quantum Coherence
  in Mesoscopic Systems}}}\ (\bibinfo  {publisher} {Springer},\ \bibinfo {year}
  {1991})\ pp.\ \bibinfo {pages} {583--595}\BibitemShut {NoStop}%
\bibitem [{\citenamefont {Stöckmann}(1999)}]{stockmann1999chaos}%
  \BibitemOpen
  \bibfield  {author} {\bibinfo {author} {\bibfnamefont {H.-J.}\ \bibnamefont
  {Stöckmann}},\ }\href {\doibase 10.1017/CBO9780511524622} {\emph {\bibinfo
  {title} {Quantum Chaos: An Introduction}}}\ (\bibinfo  {publisher} {Cambridge
  University Press},\ \bibinfo {year} {1999})\BibitemShut {NoStop}%
\bibitem [{\citenamefont {Gutzwiller}(2013)}]{gutzwiller2013chaos}%
  \BibitemOpen
  \bibfield  {author} {\bibinfo {author} {\bibfnamefont {M.~C.}\ \bibnamefont
  {Gutzwiller}},\ }\href@noop {} {\emph {\bibinfo {title} {Chaos in classical
  and quantum mechanics}}},\ Vol.~\bibinfo {volume} {1}\ (\bibinfo  {publisher}
  {Springer Science \& Business Media},\ \bibinfo {year} {2013})\BibitemShut
  {NoStop}%
\bibitem [{\citenamefont {Berry}\ \emph {et~al.}(1977)\citenamefont {Berry},
  \citenamefont {Tabor},\ and\ \citenamefont {Ziman}}]{berry1977level}%
  \BibitemOpen
  \bibfield  {author} {\bibinfo {author} {\bibfnamefont {M.~V.}\ \bibnamefont
  {Berry}}, \bibinfo {author} {\bibfnamefont {M.}~\bibnamefont {Tabor}}, \ and\
  \bibinfo {author} {\bibfnamefont {J.~M.}\ \bibnamefont {Ziman}},\ }\href
  {\doibase 10.1098/rspa.1977.0140} {\bibfield  {journal} {\bibinfo  {journal}
  {Proc. R. Soc. London, Ser. A}\ }\textbf {\bibinfo {volume} {356}},\ \bibinfo
  {pages} {375} (\bibinfo {year} {1977})}\BibitemShut {NoStop}%
\bibitem [{\citenamefont {Bohigas}\ \emph {et~al.}(1984)\citenamefont
  {Bohigas}, \citenamefont {Giannoni},\ and\ \citenamefont
  {Schmit}}]{bohigas1984characterization}%
  \BibitemOpen
  \bibfield  {author} {\bibinfo {author} {\bibfnamefont {O.}~\bibnamefont
  {Bohigas}}, \bibinfo {author} {\bibfnamefont {M.~J.}\ \bibnamefont
  {Giannoni}}, \ and\ \bibinfo {author} {\bibfnamefont {C.}~\bibnamefont
  {Schmit}},\ }\href {\doibase 10.1103/PhysRevLett.52.1} {\bibfield  {journal}
  {\bibinfo  {journal} {Phys. Rev. Lett.}\ }\textbf {\bibinfo {volume} {52}},\
  \bibinfo {pages} {1} (\bibinfo {year} {1984})}\BibitemShut {NoStop}%
\bibitem [{\citenamefont {Brody}\ \emph {et~al.}(1981)\citenamefont {Brody},
  \citenamefont {Flores}, \citenamefont {French}, \citenamefont {Mello},
  \citenamefont {Pandey},\ and\ \citenamefont {Wong}}]{brody1981random}%
  \BibitemOpen
  \bibfield  {author} {\bibinfo {author} {\bibfnamefont {T.~A.}\ \bibnamefont
  {Brody}}, \bibinfo {author} {\bibfnamefont {J.}~\bibnamefont {Flores}},
  \bibinfo {author} {\bibfnamefont {J.~B.}\ \bibnamefont {French}}, \bibinfo
  {author} {\bibfnamefont {P.~A.}\ \bibnamefont {Mello}}, \bibinfo {author}
  {\bibfnamefont {A.}~\bibnamefont {Pandey}}, \ and\ \bibinfo {author}
  {\bibfnamefont {S.~S.~M.}\ \bibnamefont {Wong}},\ }\href {\doibase
  10.1103/RevModPhys.53.385} {\bibfield  {journal} {\bibinfo  {journal} {Rev.
  Mod. Phys.}\ }\textbf {\bibinfo {volume} {53}},\ \bibinfo {pages} {385}
  (\bibinfo {year} {1981})}\BibitemShut {NoStop}%
\bibitem [{\citenamefont {Hsu}\ and\ \citenamefont
  {Angl\`es~d'Auriac}(1993)}]{hsu1993level}%
  \BibitemOpen
  \bibfield  {author} {\bibinfo {author} {\bibfnamefont {T.~C.}\ \bibnamefont
  {Hsu}}\ and\ \bibinfo {author} {\bibfnamefont {J.~C.}\ \bibnamefont
  {Angl\`es~d'Auriac}},\ }\href {\doibase 10.1103/PhysRevB.47.14291} {\bibfield
   {journal} {\bibinfo  {journal} {Phys. Rev. B}\ }\textbf {\bibinfo {volume}
  {47}},\ \bibinfo {pages} {14291} (\bibinfo {year} {1993})}\BibitemShut
  {NoStop}%
\bibitem [{\citenamefont {Zelevinsky}\ \emph {et~al.}(1996)\citenamefont
  {Zelevinsky}, \citenamefont {Brown}, \citenamefont {Frazier},\ and\
  \citenamefont {Horoi}}]{zelevinsky1996nuclear}%
  \BibitemOpen
  \bibfield  {author} {\bibinfo {author} {\bibfnamefont {V.}~\bibnamefont
  {Zelevinsky}}, \bibinfo {author} {\bibfnamefont {B.}~\bibnamefont {Brown}},
  \bibinfo {author} {\bibfnamefont {N.}~\bibnamefont {Frazier}}, \ and\
  \bibinfo {author} {\bibfnamefont {M.}~\bibnamefont {Horoi}},\ }\href
  {\doibase https://doi.org/10.1016/S0370-1573(96)00007-5} {\bibfield
  {journal} {\bibinfo  {journal} {Phys. Rep.}\ }\textbf {\bibinfo {volume}
  {276}},\ \bibinfo {pages} {85} (\bibinfo {year} {1996})}\BibitemShut
  {NoStop}%
\bibitem [{\citenamefont {Kota}(2001)}]{kota2001random}%
  \BibitemOpen
  \bibfield  {author} {\bibinfo {author} {\bibfnamefont {V.}~\bibnamefont
  {Kota}},\ }\href {\doibase https://doi.org/10.1016/S0370-1573(00)00113-7}
  {\bibfield  {journal} {\bibinfo  {journal} {Phys. Rep.}\ }\textbf {\bibinfo
  {volume} {347}},\ \bibinfo {pages} {223} (\bibinfo {year}
  {2001})}\BibitemShut {NoStop}%
\bibitem [{\citenamefont {Santos}(2004)}]{santos2004integrability}%
  \BibitemOpen
  \bibfield  {author} {\bibinfo {author} {\bibfnamefont {L.~F.}\ \bibnamefont
  {Santos}},\ }\href {\doibase 10.1088/0305-4470/37/17/004} {\bibfield
  {journal} {\bibinfo  {journal} {J. Phys. A}\ }\textbf {\bibinfo {volume}
  {37}},\ \bibinfo {pages} {4723} (\bibinfo {year} {2004})}\BibitemShut
  {NoStop}%
\bibitem [{\citenamefont {Kudo}\ and\ \citenamefont
  {Deguchi}(2005)}]{kudo2005level}%
  \BibitemOpen
  \bibfield  {author} {\bibinfo {author} {\bibfnamefont {K.}~\bibnamefont
  {Kudo}}\ and\ \bibinfo {author} {\bibfnamefont {T.}~\bibnamefont {Deguchi}},\
  }\href {\doibase 10.1143/JPSJ.74.1992} {\bibfield  {journal} {\bibinfo
  {journal} {J. Phys. Soc. Jpn.}\ }\textbf {\bibinfo {volume} {74}},\ \bibinfo
  {pages} {1992} (\bibinfo {year} {2005})}\BibitemShut {NoStop}%
\bibitem [{\citenamefont {Santos}\ and\ \citenamefont
  {Rigol}(2010)}]{santos2010onset}%
  \BibitemOpen
  \bibfield  {author} {\bibinfo {author} {\bibfnamefont {L.~F.}\ \bibnamefont
  {Santos}}\ and\ \bibinfo {author} {\bibfnamefont {M.}~\bibnamefont {Rigol}},\
  }\href {\doibase 10.1103/PhysRevE.81.036206} {\bibfield  {journal} {\bibinfo
  {journal} {Phys. Rev. E}\ }\textbf {\bibinfo {volume} {81}},\ \bibinfo
  {pages} {036206} (\bibinfo {year} {2010})}\BibitemShut {NoStop}%
\bibitem [{\citenamefont {Pal}\ and\ \citenamefont {Huse}(2010)}]{pal2010many}%
  \BibitemOpen
  \bibfield  {author} {\bibinfo {author} {\bibfnamefont {A.}~\bibnamefont
  {Pal}}\ and\ \bibinfo {author} {\bibfnamefont {D.~A.}\ \bibnamefont {Huse}},\
  }\href {\doibase 10.1103/PhysRevB.82.174411} {\bibfield  {journal} {\bibinfo
  {journal} {Phys. Rev. B}\ }\textbf {\bibinfo {volume} {82}},\ \bibinfo
  {pages} {174411} (\bibinfo {year} {2010})}\BibitemShut {NoStop}%
\bibitem [{\citenamefont {Gubin}\ and\ \citenamefont
  {F.~Santos}(2012)}]{gubin2012quantum}%
  \BibitemOpen
  \bibfield  {author} {\bibinfo {author} {\bibfnamefont {A.}~\bibnamefont
  {Gubin}}\ and\ \bibinfo {author} {\bibfnamefont {L.}~\bibnamefont
  {F.~Santos}},\ }\href {\doibase 10.1119/1.3671068} {\bibfield  {journal}
  {\bibinfo  {journal} {Am. J. Phys.}\ }\textbf {\bibinfo {volume} {80}},\
  \bibinfo {pages} {246} (\bibinfo {year} {2012})}\BibitemShut {NoStop}%
\bibitem [{\citenamefont {Abanin}\ \emph {et~al.}(2019)\citenamefont {Abanin},
  \citenamefont {Altman}, \citenamefont {Bloch},\ and\ \citenamefont
  {Serbyn}}]{abanin2019many}%
  \BibitemOpen
  \bibfield  {author} {\bibinfo {author} {\bibfnamefont {D.~A.}\ \bibnamefont
  {Abanin}}, \bibinfo {author} {\bibfnamefont {E.}~\bibnamefont {Altman}},
  \bibinfo {author} {\bibfnamefont {I.}~\bibnamefont {Bloch}}, \ and\ \bibinfo
  {author} {\bibfnamefont {M.}~\bibnamefont {Serbyn}},\ }\href {\doibase
  10.1103/RevModPhys.91.021001} {\bibfield  {journal} {\bibinfo  {journal}
  {Rev. Mod. Phys.}\ }\textbf {\bibinfo {volume} {91}},\ \bibinfo {pages}
  {021001} (\bibinfo {year} {2019})}\BibitemShut {NoStop}%
\bibitem [{\citenamefont {Breuer}\ and\ \citenamefont
  {Petruccione}(2002)}]{breuer2002theory}%
  \BibitemOpen
  \bibfield  {author} {\bibinfo {author} {\bibfnamefont {H.-P.}\ \bibnamefont
  {Breuer}}\ and\ \bibinfo {author} {\bibfnamefont {F.}~\bibnamefont
  {Petruccione}},\ }\href@noop {} {\emph {\bibinfo {title} {The theory of open
  quantum systems}}}\ (\bibinfo  {publisher} {Oxford University Press},\
  \bibinfo {year} {2002})\BibitemShut {NoStop}%
\bibitem [{\citenamefont {Grobe}\ \emph {et~al.}(1988)\citenamefont {Grobe},
  \citenamefont {Haake},\ and\ \citenamefont {Sommers}}]{grobe1988quantum}%
  \BibitemOpen
  \bibfield  {author} {\bibinfo {author} {\bibfnamefont {R.}~\bibnamefont
  {Grobe}}, \bibinfo {author} {\bibfnamefont {F.}~\bibnamefont {Haake}}, \ and\
  \bibinfo {author} {\bibfnamefont {H.-J.}\ \bibnamefont {Sommers}},\ }\href
  {\doibase 10.1103/PhysRevLett.61.1899} {\bibfield  {journal} {\bibinfo
  {journal} {Phys. Rev. Lett.}\ }\textbf {\bibinfo {volume} {61}},\ \bibinfo
  {pages} {1899} (\bibinfo {year} {1988})}\BibitemShut {NoStop}%
\bibitem [{\citenamefont {Ginibre}(1965)}]{ginibre1965statistical}%
  \BibitemOpen
  \bibfield  {author} {\bibinfo {author} {\bibfnamefont {J.}~\bibnamefont
  {Ginibre}},\ }\href {\doibase 10.1063/1.1704292} {\bibfield  {journal}
  {\bibinfo  {journal} {J. Math. Phys. (N.Y.)}\ }\textbf {\bibinfo {volume}
  {6}},\ \bibinfo {pages} {440} (\bibinfo {year} {1965})}\BibitemShut {NoStop}%
\bibitem [{\citenamefont {Forrester}(2010)}]{forrester2010log}%
  \BibitemOpen
  \bibfield  {author} {\bibinfo {author} {\bibfnamefont {P.~J.}\ \bibnamefont
  {Forrester}},\ }\href@noop {} {\emph {\bibinfo {title} {Log-Gases and Random
  Matrices}}}\ (\bibinfo  {publisher} {Princeton University Press},\ \bibinfo
  {year} {2010})\BibitemShut {NoStop}%
\bibitem [{\citenamefont {Hamazaki}\ \emph {et~al.}(2019)\citenamefont
  {Hamazaki}, \citenamefont {Kawabata},\ and\ \citenamefont
  {Ueda}}]{hamazaki2019non}%
  \BibitemOpen
  \bibfield  {author} {\bibinfo {author} {\bibfnamefont {R.}~\bibnamefont
  {Hamazaki}}, \bibinfo {author} {\bibfnamefont {K.}~\bibnamefont {Kawabata}},
  \ and\ \bibinfo {author} {\bibfnamefont {M.}~\bibnamefont {Ueda}},\ }\href
  {\doibase 10.1103/PhysRevLett.123.090603} {\bibfield  {journal} {\bibinfo
  {journal} {Phys. Rev. Lett.}\ }\textbf {\bibinfo {volume} {123}},\ \bibinfo
  {pages} {090603} (\bibinfo {year} {2019})}\BibitemShut {NoStop}%
\bibitem [{\citenamefont {Akemann}\ \emph {et~al.}(2019)\citenamefont
  {Akemann}, \citenamefont {Kieburg}, \citenamefont {Mielke},\ and\
  \citenamefont {Prosen}}]{akemann2019universal}%
  \BibitemOpen
  \bibfield  {author} {\bibinfo {author} {\bibfnamefont {G.}~\bibnamefont
  {Akemann}}, \bibinfo {author} {\bibfnamefont {M.}~\bibnamefont {Kieburg}},
  \bibinfo {author} {\bibfnamefont {A.}~\bibnamefont {Mielke}}, \ and\ \bibinfo
  {author} {\bibfnamefont {T.}~\bibnamefont {Prosen}},\ }\href {\doibase
  10.1103/PhysRevLett.123.254101} {\bibfield  {journal} {\bibinfo  {journal}
  {Phys. Rev. Lett.}\ }\textbf {\bibinfo {volume} {123}},\ \bibinfo {pages}
  {254101} (\bibinfo {year} {2019})}\BibitemShut {NoStop}%
\bibitem [{\citenamefont {Dicke}(1954)}]{dicke1954coherence}%
  \BibitemOpen
  \bibfield  {author} {\bibinfo {author} {\bibfnamefont {R.~H.}\ \bibnamefont
  {Dicke}},\ }\href {\doibase 10.1103/PhysRev.93.99} {\bibfield  {journal}
  {\bibinfo  {journal} {Phys. Rev.}\ }\textbf {\bibinfo {volume} {93}},\
  \bibinfo {pages} {99} (\bibinfo {year} {1954})}\BibitemShut {NoStop}%
\bibitem [{\citenamefont {Gross}\ and\ \citenamefont
  {Haroche}(1982)}]{gross1982superradiance}%
  \BibitemOpen
  \bibfield  {author} {\bibinfo {author} {\bibfnamefont {M.}~\bibnamefont
  {Gross}}\ and\ \bibinfo {author} {\bibfnamefont {S.}~\bibnamefont
  {Haroche}},\ }\href {\doibase https://doi.org/10.1016/0370-1573(82)90102-8}
  {\bibfield  {journal} {\bibinfo  {journal} {Phys. Rep.}\ }\textbf {\bibinfo
  {volume} {93}},\ \bibinfo {pages} {301} (\bibinfo {year} {1982})}\BibitemShut
  {NoStop}%
\bibitem [{\citenamefont {Garraway}(2011)}]{garraway2011dicke}%
  \BibitemOpen
  \bibfield  {author} {\bibinfo {author} {\bibfnamefont {B.~M.}\ \bibnamefont
  {Garraway}},\ }\href {\doibase 10.1098/rsta.2010.0333} {\bibfield  {journal}
  {\bibinfo  {journal} {Philos. Trans. R. Soc. London, Ser. A}\ }\textbf
  {\bibinfo {volume} {369}},\ \bibinfo {pages} {1137} (\bibinfo {year}
  {2011})}\BibitemShut {NoStop}%
\bibitem [{\citenamefont {Hepp}\ and\ \citenamefont
  {Lieb}(1973)}]{hepp1973superradiant}%
  \BibitemOpen
  \bibfield  {author} {\bibinfo {author} {\bibfnamefont {K.}~\bibnamefont
  {Hepp}}\ and\ \bibinfo {author} {\bibfnamefont {E.~H.}\ \bibnamefont
  {Lieb}},\ }\href {\doibase https://doi.org/10.1016/0003-4916(73)90039-0}
  {\bibfield  {journal} {\bibinfo  {journal} {Ann. Phys. (N.Y.)}\ }\textbf
  {\bibinfo {volume} {76}},\ \bibinfo {pages} {360} (\bibinfo {year}
  {1973})}\BibitemShut {NoStop}%
\bibitem [{\citenamefont {Kirton}\ \emph {et~al.}(2019)\citenamefont {Kirton},
  \citenamefont {Roses}, \citenamefont {Keeling},\ and\ \citenamefont
  {Dalla~Torre}}]{kirton2019introduction}%
  \BibitemOpen
  \bibfield  {author} {\bibinfo {author} {\bibfnamefont {P.}~\bibnamefont
  {Kirton}}, \bibinfo {author} {\bibfnamefont {M.~M.}\ \bibnamefont {Roses}},
  \bibinfo {author} {\bibfnamefont {J.}~\bibnamefont {Keeling}}, \ and\
  \bibinfo {author} {\bibfnamefont {E.~G.}\ \bibnamefont {Dalla~Torre}},\
  }\href {\doibase https://doi.org/10.1002/qute.201800043} {\bibfield
  {journal} {\bibinfo  {journal} {Advanced Quantum Technologies}\ }\textbf
  {\bibinfo {volume} {2}},\ \bibinfo {pages} {1800043} (\bibinfo {year}
  {2019})}\BibitemShut {NoStop}%
\bibitem [{\citenamefont {Emary}\ and\ \citenamefont
  {Brandes}(2003{\natexlab{a}})}]{emary2003chaos1}%
  \BibitemOpen
  \bibfield  {author} {\bibinfo {author} {\bibfnamefont {C.}~\bibnamefont
  {Emary}}\ and\ \bibinfo {author} {\bibfnamefont {T.}~\bibnamefont
  {Brandes}},\ }\href {\doibase 10.1103/PhysRevLett.90.044101} {\bibfield
  {journal} {\bibinfo  {journal} {Phys. Rev. Lett.}\ }\textbf {\bibinfo
  {volume} {90}},\ \bibinfo {pages} {044101} (\bibinfo {year}
  {2003}{\natexlab{a}})}\BibitemShut {NoStop}%
\bibitem [{\citenamefont {Emary}\ and\ \citenamefont
  {Brandes}(2003{\natexlab{b}})}]{emary2003chaos2}%
  \BibitemOpen
  \bibfield  {author} {\bibinfo {author} {\bibfnamefont {C.}~\bibnamefont
  {Emary}}\ and\ \bibinfo {author} {\bibfnamefont {T.}~\bibnamefont
  {Brandes}},\ }\href {\doibase 10.1103/PhysRevE.67.066203} {\bibfield
  {journal} {\bibinfo  {journal} {Phys. Rev. E}\ }\textbf {\bibinfo {volume}
  {67}},\ \bibinfo {pages} {066203} (\bibinfo {year}
  {2003}{\natexlab{b}})}\BibitemShut {NoStop}%
\bibitem [{\citenamefont {Albert}\ and\ \citenamefont
  {Jiang}(2014)}]{albert2014symmetries}%
  \BibitemOpen
  \bibfield  {author} {\bibinfo {author} {\bibfnamefont {V.~V.}\ \bibnamefont
  {Albert}}\ and\ \bibinfo {author} {\bibfnamefont {L.}~\bibnamefont {Jiang}},\
  }\href {\doibase 10.1103/PhysRevA.89.022118} {\bibfield  {journal} {\bibinfo
  {journal} {Phys. Rev. A}\ }\textbf {\bibinfo {volume} {89}},\ \bibinfo
  {pages} {022118} (\bibinfo {year} {2014})}\BibitemShut {NoStop}%
\bibitem [{\citenamefont {Bu\v{c}a}\ and\ \citenamefont
  {Prosen}(2012)}]{buvca2012note}%
  \BibitemOpen
  \bibfield  {author} {\bibinfo {author} {\bibfnamefont {B.}~\bibnamefont
  {Bu\v{c}a}}\ and\ \bibinfo {author} {\bibfnamefont {T.}~\bibnamefont
  {Prosen}},\ }\href {\doibase 10.1088/1367-2630/14/7/073007} {\bibfield
  {journal} {\bibinfo  {journal} {New J. Phys.}\ }\textbf {\bibinfo {volume}
  {14}},\ \bibinfo {pages} {073007} (\bibinfo {year} {2012})}\BibitemShut
  {NoStop}%
\bibitem [{\citenamefont {Lieu}\ \emph {et~al.}(2020)\citenamefont {Lieu},
  \citenamefont {Belyansky}, \citenamefont {Young}, \citenamefont {Lundgren},
  \citenamefont {Albert},\ and\ \citenamefont {Gorshkov}}]{lieu2020symmetry}%
  \BibitemOpen
  \bibfield  {author} {\bibinfo {author} {\bibfnamefont {S.}~\bibnamefont
  {Lieu}}, \bibinfo {author} {\bibfnamefont {R.}~\bibnamefont {Belyansky}},
  \bibinfo {author} {\bibfnamefont {J.~T.}\ \bibnamefont {Young}}, \bibinfo
  {author} {\bibfnamefont {R.}~\bibnamefont {Lundgren}}, \bibinfo {author}
  {\bibfnamefont {V.~V.}\ \bibnamefont {Albert}}, \ and\ \bibinfo {author}
  {\bibfnamefont {A.~V.}\ \bibnamefont {Gorshkov}},\ }\href {\doibase
  10.1103/PhysRevLett.125.240405} {\bibfield  {journal} {\bibinfo  {journal}
  {Phys. Rev. Lett.}\ }\textbf {\bibinfo {volume} {125}},\ \bibinfo {pages}
  {240405} (\bibinfo {year} {2020})}\BibitemShut {NoStop}%
\bibitem [{SM()}]{SM}%
  \BibitemOpen
  \href@noop {} {\bibinfo  {journal} {Supplementary material}\ }\BibitemShut
  {NoStop}%
\bibitem [{\citenamefont {Dimer}\ \emph {et~al.}(2007)\citenamefont {Dimer},
  \citenamefont {Estienne}, \citenamefont {Parkins},\ and\ \citenamefont
  {Carmichael}}]{Dimer2007Dicke}%
  \BibitemOpen
\bibfield  {journal} {  }\bibfield  {author} {\bibinfo {author} {\bibfnamefont
  {F.}~\bibnamefont {Dimer}}, \bibinfo {author} {\bibfnamefont
  {B.}~\bibnamefont {Estienne}}, \bibinfo {author} {\bibfnamefont {A.~S.}\
  \bibnamefont {Parkins}}, \ and\ \bibinfo {author} {\bibfnamefont {H.~J.}\
  \bibnamefont {Carmichael}},\ }\href {\doibase 10.1103/PhysRevA.75.013804}
  {\bibfield  {journal} {\bibinfo  {journal} {Phys. Rev. A}\ }\textbf {\bibinfo
  {volume} {75}},\ \bibinfo {pages} {013804} (\bibinfo {year}
  {2007})}\BibitemShut {NoStop}%
\bibitem [{\citenamefont {Roses}\ and\ \citenamefont
  {Dalla~Torre}(2020)}]{roses2020dicke}%
  \BibitemOpen
  \bibfield  {author} {\bibinfo {author} {\bibfnamefont {M.~M.}\ \bibnamefont
  {Roses}}\ and\ \bibinfo {author} {\bibfnamefont {E.~G.}\ \bibnamefont
  {Dalla~Torre}},\ }\href {\doibase
  https://doi.org/10.1371/journal.pone.0235197} {\bibfield  {journal} {\bibinfo
   {journal} {PLoS ONE}\ }\textbf {\bibinfo {volume} {15}},\ \bibinfo {pages}
  {e0235197} (\bibinfo {year} {2020})}\BibitemShut {NoStop}%
\bibitem [{\citenamefont {Rosenzweig}\ and\ \citenamefont
  {Porter}(1960)}]{rosenzweig1960repulsion}%
  \BibitemOpen
  \bibfield  {author} {\bibinfo {author} {\bibfnamefont {N.}~\bibnamefont
  {Rosenzweig}}\ and\ \bibinfo {author} {\bibfnamefont {C.~E.}\ \bibnamefont
  {Porter}},\ }\href {\doibase 10.1103/PhysRev.120.1698} {\bibfield  {journal}
  {\bibinfo  {journal} {Phys. Rev.}\ }\textbf {\bibinfo {volume} {120}},\
  \bibinfo {pages} {1698} (\bibinfo {year} {1960})}\BibitemShut {NoStop}%
\bibitem [{\citenamefont {S\'a}\ \emph {et~al.}(2020)\citenamefont {S\'a},
  \citenamefont {Ribeiro},\ and\ \citenamefont {Prosen}}]{sa2020complex}%
  \BibitemOpen
  \bibfield  {author} {\bibinfo {author} {\bibfnamefont {L.}~\bibnamefont
  {S\'a}}, \bibinfo {author} {\bibfnamefont {P.}~\bibnamefont {Ribeiro}}, \
  and\ \bibinfo {author} {\bibfnamefont {T.}~\bibnamefont {Prosen}},\ }\href
  {\doibase 10.1103/PhysRevX.10.021019} {\bibfield  {journal} {\bibinfo
  {journal} {Phys. Rev. X}\ }\textbf {\bibinfo {volume} {10}},\ \bibinfo
  {pages} {021019} (\bibinfo {year} {2020})}\BibitemShut {NoStop}%
\bibitem [{\citenamefont {Oganesyan}\ and\ \citenamefont
  {Huse}(2007)}]{oganesyan2007localization}%
  \BibitemOpen
  \bibfield  {author} {\bibinfo {author} {\bibfnamefont {V.}~\bibnamefont
  {Oganesyan}}\ and\ \bibinfo {author} {\bibfnamefont {D.~A.}\ \bibnamefont
  {Huse}},\ }\href {\doibase 10.1103/PhysRevB.75.155111} {\bibfield  {journal}
  {\bibinfo  {journal} {Phys. Rev. B}\ }\textbf {\bibinfo {volume} {75}},\
  \bibinfo {pages} {155111} (\bibinfo {year} {2007})}\BibitemShut {NoStop}%
\bibitem [{\citenamefont {Atas}\ \emph {et~al.}(2013)\citenamefont {Atas},
  \citenamefont {Bogomolny}, \citenamefont {Giraud},\ and\ \citenamefont
  {Roux}}]{atas2013distribution}%
  \BibitemOpen
  \bibfield  {author} {\bibinfo {author} {\bibfnamefont {Y.~Y.}\ \bibnamefont
  {Atas}}, \bibinfo {author} {\bibfnamefont {E.}~\bibnamefont {Bogomolny}},
  \bibinfo {author} {\bibfnamefont {O.}~\bibnamefont {Giraud}}, \ and\ \bibinfo
  {author} {\bibfnamefont {G.}~\bibnamefont {Roux}},\ }\href {\doibase
  10.1103/PhysRevLett.110.084101} {\bibfield  {journal} {\bibinfo  {journal}
  {Phys. Rev. Lett.}\ }\textbf {\bibinfo {volume} {110}},\ \bibinfo {pages}
  {084101} (\bibinfo {year} {2013})}\BibitemShut {NoStop}%
\end{thebibliography}%


\begin{thebibliography}{9}%
\makeatletter
\providecommand \@ifxundefined [1]{%
 \@ifx{#1\undefined}
}%
\providecommand \@ifnum [1]{%
 \ifnum #1\expandafter \@firstoftwo
 \else \expandafter \@secondoftwo
 \fi
}%
\providecommand \@ifx [1]{%
 \ifx #1\expandafter \@firstoftwo
 \else \expandafter \@secondoftwo
 \fi
}%
\providecommand \natexlab [1]{#1}%
\providecommand \enquote  [1]{``#1''}%
\providecommand \bibnamefont  [1]{#1}%
\providecommand \bibfnamefont [1]{#1}%
\providecommand \citenamefont [1]{#1}%
\providecommand \href@noop [0]{\@secondoftwo}%
\providecommand \href [0]{\begingroup \@sanitize@url \@href}%
\providecommand \@href[1]{\@@startlink{#1}\@@href}%
\providecommand \@@href[1]{\endgroup#1\@@endlink}%
\providecommand \@sanitize@url [0]{\catcode `\\12\catcode `\$12\catcode
  `\&12\catcode `\#12\catcode `\^12\catcode `\_12\catcode `\%12\relax}%
\providecommand \@@startlink[1]{}%
\providecommand \@@endlink[0]{}%
\providecommand \url  [0]{\begingroup\@sanitize@url \@url }%
\providecommand \@url [1]{\endgroup\@href {#1}{\urlprefix }}%
\providecommand \urlprefix  [0]{URL }%
\providecommand \Eprint [0]{\href }%
\providecommand \doibase [0]{http://dx.doi.org/}%
\providecommand \selectlanguage [0]{\@gobble}%
\providecommand \bibinfo  [0]{\@secondoftwo}%
\providecommand \bibfield  [0]{\@secondoftwo}%
\providecommand \translation [1]{[#1]}%
\providecommand \BibitemOpen [0]{}%
\providecommand \bibitemStop [0]{}%
\providecommand \bibitemNoStop [0]{.\EOS\space}%
\providecommand \EOS [0]{\spacefactor3000\relax}%
\providecommand \BibitemShut  [1]{\csname bibitem#1\endcsname}%
\let\auto@bib@innerbib\@empty
\bibitem [{\citenamefont {Mukamel}(1999)}]{mukamel1999principles}%
  \BibitemOpen
  \bibfield  {author} {\bibinfo {author} {\bibfnamefont {S.}~\bibnamefont
  {Mukamel}},\ }\href@noop {} {\emph {\bibinfo {title} {Principles of nonlinear
  optical spectroscopy}}},\ \bibinfo {number} {6}\ (\bibinfo  {publisher}
  {Oxford University Press},\ \bibinfo {year} {1999})\BibitemShut {NoStop}%
\bibitem [{\citenamefont {Haake}(1991)}]{haake1991quantum}%
  \BibitemOpen
  \bibfield  {author} {\bibinfo {author} {\bibfnamefont {F.}~\bibnamefont
  {Haake}},\ }in\ \href@noop {} {\emph {\bibinfo {booktitle} {Quantum Coherence
  in Mesoscopic Systems}}}\ (\bibinfo  {publisher} {Springer},\ \bibinfo {year}
  {1991})\ pp.\ \bibinfo {pages} {583--595}\BibitemShut {NoStop}%
\bibitem [{\citenamefont {Albert}\ and\ \citenamefont
  {Jiang}(2014)}]{albert2014symmetries}%
  \BibitemOpen
  \bibfield  {author} {\bibinfo {author} {\bibfnamefont {V.~V.}\ \bibnamefont
  {Albert}}\ and\ \bibinfo {author} {\bibfnamefont {L.}~\bibnamefont {Jiang}},\
  }\href {\doibase 10.1103/PhysRevA.89.022118} {\bibfield  {journal} {\bibinfo
  {journal} {Phys. Rev. A}\ }\textbf {\bibinfo {volume} {89}},\ \bibinfo
  {pages} {022118} (\bibinfo {year} {2014})}\BibitemShut {NoStop}%
\bibitem [{\citenamefont {Bu\v{c}a}\ and\ \citenamefont
  {Prosen}(2012)}]{buvca2012note}%
  \BibitemOpen
  \bibfield  {author} {\bibinfo {author} {\bibfnamefont {B.}~\bibnamefont
  {Bu\v{c}a}}\ and\ \bibinfo {author} {\bibfnamefont {T.}~\bibnamefont
  {Prosen}},\ }\href {\doibase 10.1088/1367-2630/14/7/073007} {\bibfield
  {journal} {\bibinfo  {journal} {New J. Phys.}\ }\textbf {\bibinfo {volume}
  {14}},\ \bibinfo {pages} {073007} (\bibinfo {year} {2012})}\BibitemShut
  {NoStop}%
\bibitem [{\citenamefont {Lieu}\ \emph {et~al.}(2020)\citenamefont {Lieu},
  \citenamefont {Belyansky}, \citenamefont {Young}, \citenamefont {Lundgren},
  \citenamefont {Albert},\ and\ \citenamefont {Gorshkov}}]{lieu2020symmetry}%
  \BibitemOpen
  \bibfield  {author} {\bibinfo {author} {\bibfnamefont {S.}~\bibnamefont
  {Lieu}}, \bibinfo {author} {\bibfnamefont {R.}~\bibnamefont {Belyansky}},
  \bibinfo {author} {\bibfnamefont {J.~T.}\ \bibnamefont {Young}}, \bibinfo
  {author} {\bibfnamefont {R.}~\bibnamefont {Lundgren}}, \bibinfo {author}
  {\bibfnamefont {V.~V.}\ \bibnamefont {Albert}}, \ and\ \bibinfo {author}
  {\bibfnamefont {A.~V.}\ \bibnamefont {Gorshkov}},\ }\href {\doibase
  10.1103/PhysRevLett.125.240405} {\bibfield  {journal} {\bibinfo  {journal}
  {Phys. Rev. Lett.}\ }\textbf {\bibinfo {volume} {125}},\ \bibinfo {pages}
  {240405} (\bibinfo {year} {2020})}\BibitemShut {NoStop}%
\bibitem [{\citenamefont {Markum}\ \emph {et~al.}(1999)\citenamefont {Markum},
  \citenamefont {Pullirsch},\ and\ \citenamefont {Wettig}}]{markum1999non}%
  \BibitemOpen
  \bibfield  {author} {\bibinfo {author} {\bibfnamefont {H.}~\bibnamefont
  {Markum}}, \bibinfo {author} {\bibfnamefont {R.}~\bibnamefont {Pullirsch}}, \
  and\ \bibinfo {author} {\bibfnamefont {T.}~\bibnamefont {Wettig}},\ }\href
  {\doibase 10.1103/PhysRevLett.83.484} {\bibfield  {journal} {\bibinfo
  {journal} {Phys. Rev. Lett.}\ }\textbf {\bibinfo {volume} {83}},\ \bibinfo
  {pages} {484} (\bibinfo {year} {1999})}\BibitemShut {NoStop}%
\bibitem [{\citenamefont {Akemann}\ \emph {et~al.}(2019)\citenamefont
  {Akemann}, \citenamefont {Kieburg}, \citenamefont {Mielke},\ and\
  \citenamefont {Prosen}}]{akemann2019universal}%
  \BibitemOpen
  \bibfield  {author} {\bibinfo {author} {\bibfnamefont {G.}~\bibnamefont
  {Akemann}}, \bibinfo {author} {\bibfnamefont {M.}~\bibnamefont {Kieburg}},
  \bibinfo {author} {\bibfnamefont {A.}~\bibnamefont {Mielke}}, \ and\ \bibinfo
  {author} {\bibfnamefont {T.}~\bibnamefont {Prosen}},\ }\href {\doibase
  10.1103/PhysRevLett.123.254101} {\bibfield  {journal} {\bibinfo  {journal}
  {Phys. Rev. Lett.}\ }\textbf {\bibinfo {volume} {123}},\ \bibinfo {pages}
  {254101} (\bibinfo {year} {2019})}\BibitemShut {NoStop}%
\bibitem [{\citenamefont {Hamazaki}\ \emph {et~al.}(2020)\citenamefont
  {Hamazaki}, \citenamefont {Kawabata}, \citenamefont {Kura},\ and\
  \citenamefont {Ueda}}]{hamazaki2020universality}%
  \BibitemOpen
  \bibfield  {author} {\bibinfo {author} {\bibfnamefont {R.}~\bibnamefont
  {Hamazaki}}, \bibinfo {author} {\bibfnamefont {K.}~\bibnamefont {Kawabata}},
  \bibinfo {author} {\bibfnamefont {N.}~\bibnamefont {Kura}}, \ and\ \bibinfo
  {author} {\bibfnamefont {M.}~\bibnamefont {Ueda}},\ }\href {\doibase
  10.1103/PhysRevResearch.2.023286} {\bibfield  {journal} {\bibinfo  {journal}
  {Phys. Rev. Research}\ }\textbf {\bibinfo {volume} {2}},\ \bibinfo {pages}
  {023286} (\bibinfo {year} {2020})}\BibitemShut {NoStop}%
\bibitem [{\citenamefont {S\'a}\ \emph {et~al.}(2020)\citenamefont {S\'a},
  \citenamefont {Ribeiro},\ and\ \citenamefont {Prosen}}]{sa2020complex}%
  \BibitemOpen
  \bibfield  {author} {\bibinfo {author} {\bibfnamefont {L.}~\bibnamefont
  {S\'a}}, \bibinfo {author} {\bibfnamefont {P.}~\bibnamefont {Ribeiro}}, \
  and\ \bibinfo {author} {\bibfnamefont {T.}~\bibnamefont {Prosen}},\ }\href
  {\doibase 10.1103/PhysRevX.10.021019} {\bibfield  {journal} {\bibinfo
  {journal} {Phys. Rev. X}\ }\textbf {\bibinfo {volume} {10}},\ \bibinfo
  {pages} {021019} (\bibinfo {year} {2020})}\BibitemShut {NoStop}%
\end{thebibliography}%

\end{document}


\newcommand{\titlename}{\underline{Supplementary material}\\ \medskip Dissipative quantum dynamics, phase transitions and non-Hermitian random matrices}
\title{\titlename}
\author{Mahaveer Prasad}
\email{mahaveer.prasad@icts.res.in}
 \affiliation{International Centre for Theoretical Sciences, Tata Institute of 
Fundamental Research, 560089 Bangalore, India}
\author{Hari Kumar Yadalam}
 \email{hari.kumar@icts.res.in}
\affiliation{International Centre for Theoretical Sciences, Tata Institute of 
Fundamental Research, 560089 Bangalore, India}
\affiliation{Laboratoire de Physique, \'Ecole Normale Sup\'erieure, CNRS, 
Universit\'e PSL, Sorbonne Universit\'e Universit\'e de Paris, 75005 Paris, France}%
\author{Camille Aron}
 \email{aron@ens.fr}
\affiliation{Laboratoire de Physique, \'Ecole Normale Sup\'erieure, CNRS, 
Universit\'e PSL, Sorbonne Universit\'e Universit\'e de Paris, 75005 Paris, France}%
\author{Manas Kulkarni}
 \email{manas.kulkarni@icts.res.in}
\affiliation{International Centre for Theoretical Sciences, Tata Institute of 
Fundamental Research, 560089 Bangalore, India}

\date{\today}

\maketitle


\vspace{0.8cm}


\section{Computing the spectrum of the dissipative Dicke Liouvillian}
\label{app_sec_algo}
In this Section we discuss the steps used to compute the spectrum of the Liouvillian $\mathcal{L}$ of the dissipative Dicke model introduced in Eq.~(1) of the main text. 

Let us first introduce a convenient basis of the Liouville space.
We choose to work with the basis states
\begin{align}
|\alpha \rangle \rangle  \equiv \big{|} | n_{l},m_{l}\rangle \langle n_{r},m_{r}| \, \big{\rangle} \,,
\end{align}
where $| n,m\rangle $ are the Fock states of the Dicke Hamiltonian with $n \in \mathbb{N}$ cavity excitations and $m = -S/2, -S/2 +1 , \ldots, S/2$ is the quantum number associated with the $z$-component of the spin.
We introduced the notation $| \alpha \rangle \rangle$ where the label $\alpha$ collects all the quantum numbers $n_l,m_l,n_r,m_r$. This choice of notation expresses the fact that operators on the Hilbert space are states in the Liouville space. In practice, we truncate the Hilbert and Liouville space by  introducing a cavity cutoff such that $n = 0, 1, \ldots, n_{\rm cutoff}$.

In the basis introduced above, the Liouvillian can be represented by a matrix $L$ with the elements
\begin{align}
L_{\alpha\alpha'} = \langle\langle \alpha | \mathcal{L} | \alpha' \rangle\rangle \,,
\end{align}
where the Hilbert-Schmidt inner product~\cite{mukamel1999principles, haake1991quantum} is given by 
\begin{align}
\langle\langle \alpha | \alpha ' \rangle \rangle \equiv \mathrm{Tr} \! \left[ \left( | n_{l},m_{l}\rangle \langle n_{r},m_{r}| \right)^\dagger \!  | n'_{l},m'_{l}\rangle \langle n'_{r},m'_{r}| \right]
\end{align}
and the trace is performed on the Hilbert space.

Let us recall that $\mathcal{L}$ has a weak $\mathbb{Z}_2$ symmetry~\cite{albert2014symmetries,buvca2012note,lieu2020symmetry}, namely parity:
\begin{align}
[\mathcal{L},\varPi] = 0\,,
\end{align}
where the superoperator $\varPi$ acts on the basis states as
\begin{align}
\varPi | \alpha \rangle\rangle = \varPi | n_{l},m_{l}\rangle \langle n_{r},m_{r}| = \zeta \, | \alpha \rangle\rangle \,,
\end{align}
with  $\zeta=+1$ if the number $(n_l+m_l)-(n_r+m_r)$ is even and $\zeta=-1$ if it is odd. This weak $\mathbb{Z}_2$ symmetry of $\mathcal{L}$ guarantees that it does not couple states of the Liouville space with different parities. Hence, $L_{\alpha\alpha'}$ can be organized as  a two-by-two block diagonal matrix.

To properly extract the statistical properties of the Liouvillian spectrum, we only consider its even-parity block. This avoids possible spurious overlaps of eigenvalues from the odd-parity block.
Finally, the even-parity block matrix is fed to a diagonalization algorithm adapted to complex non-Hermitian matrices using suitable numerical routines in the LAPACK library. 

\begin{figure}[h]
\centering
\includegraphics[width=\linewidth]{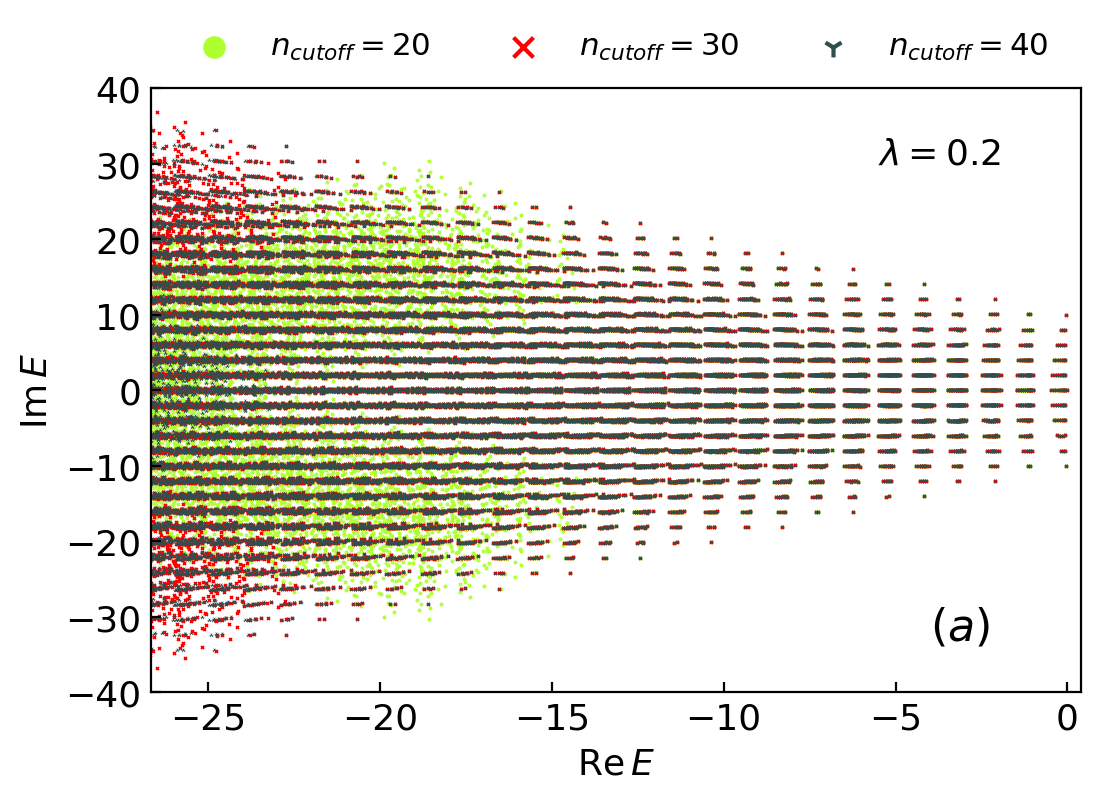}
\includegraphics[width=\linewidth]{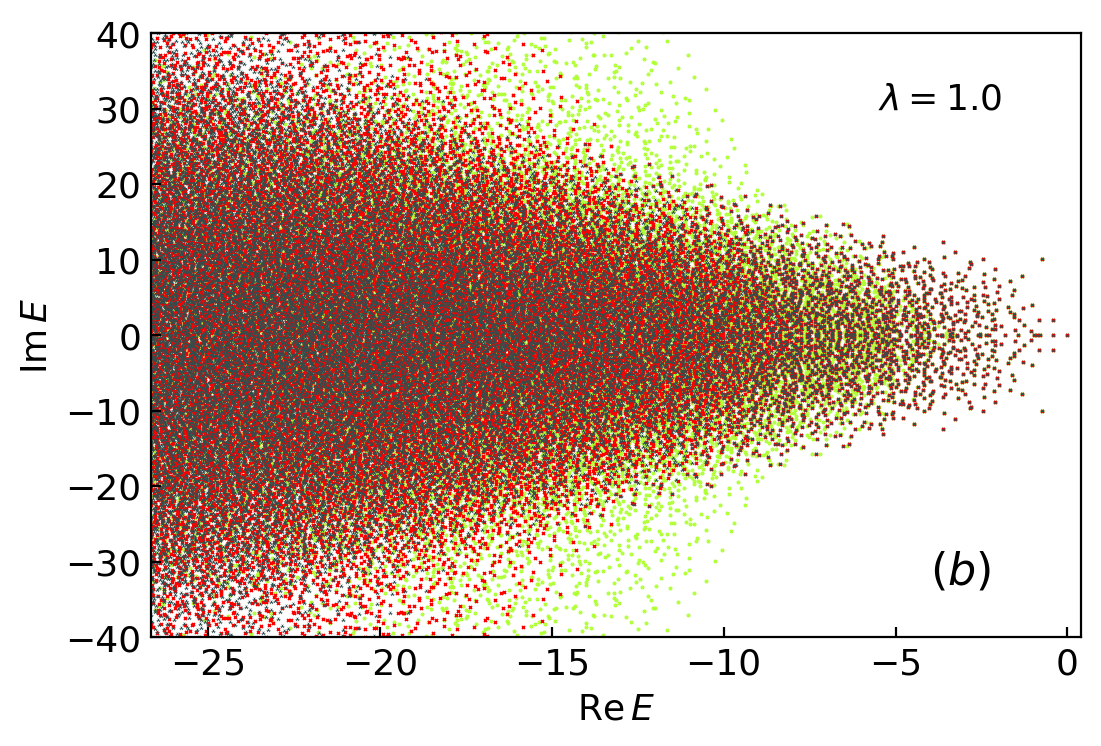}
\caption{Scattter plot of the spectrum of the Liouvillian of the dissipative Dicke model for  different $n_{\rm cutoff}=20,30,40$ in (a) the normal phase, $\lambda=0.2$ and (b) the superradiant phase, $\lambda=1.0$ ($S=10$).} 
\label{Ncut_spectrum}
\end{figure}

\section{Unfolding the complex spectrum}
In order to extract the universal features of the level-spacing statistics from the spectrum, we first perform an unfolding procedure to eliminate the system-specific features.
Many methods have been proposed to perform the unfolding in the case of a complex spectrum~\cite{markum1999non,akemann2019universal,hamazaki2020universality}. We use the method put forward in Ref.~\onlinecite{akemann2019universal} which we briefly describe below.
First, we compute the Euclidean distance of each of the $N$ complex eigenvalues to its nearest neighbor (NN)
\begin{align}
\label{eq:si}
s_i = | E_i - E_i^{\rm NN} |\,.
\end{align}
Next, we rescale these distances as
\begin{align} \label{eq:ansatz}
s_i \to s_i' = s_i \, \frac{\sqrt{\rho_{\rm av}(E_i)}}{\bar s} \,,
\end{align}
where 
 $\rho_{\rm av}(E_i)$ is the local average density approximated by 
\begin{equation}
\rho_{avg}(E) =\frac{1}{2 \pi \sigma^{2}N} \sum_{i=1}^{N} 
\exp \left( -\frac{|E-E_{i}|^2}{2 \sigma^{2}} \right) \,.
\end{equation}
We choose $\sigma$ to be greater than the global mean level spacing given by $\tilde{s} = (1/N) \sum_{i=1}^{N} s_i $. This guarantees a smooth distribution function on the scale of $\tilde{s}$.  In practice, we work with $\sigma = 4.5 \times \tilde{s}$. 
$\bar s$ in Eq.~(\ref{eq:ansatz}) is chosen to ensure that the global mean level spacing of the final distribution of the $s'_i$ is set to unity: $ (1/N) \sum_{i=1}^{N} s_i' = 1$. Finally, the statistics of nearest-level spacings are computed from $s'_i$. In the main text we drop the prime notation in $s'_i$ for the sake of simplicity.

\begin{figure}[t!]
  \includegraphics[scale=0.28]{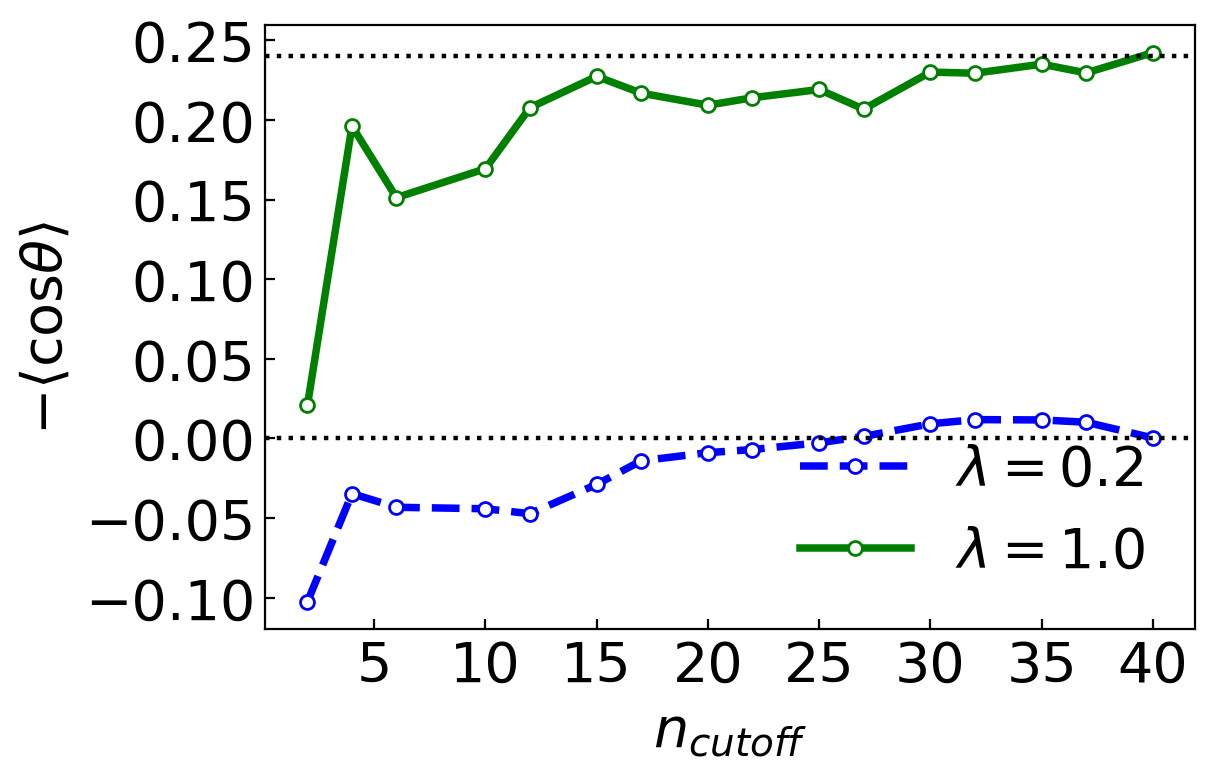}
  \includegraphics[scale=0.28]{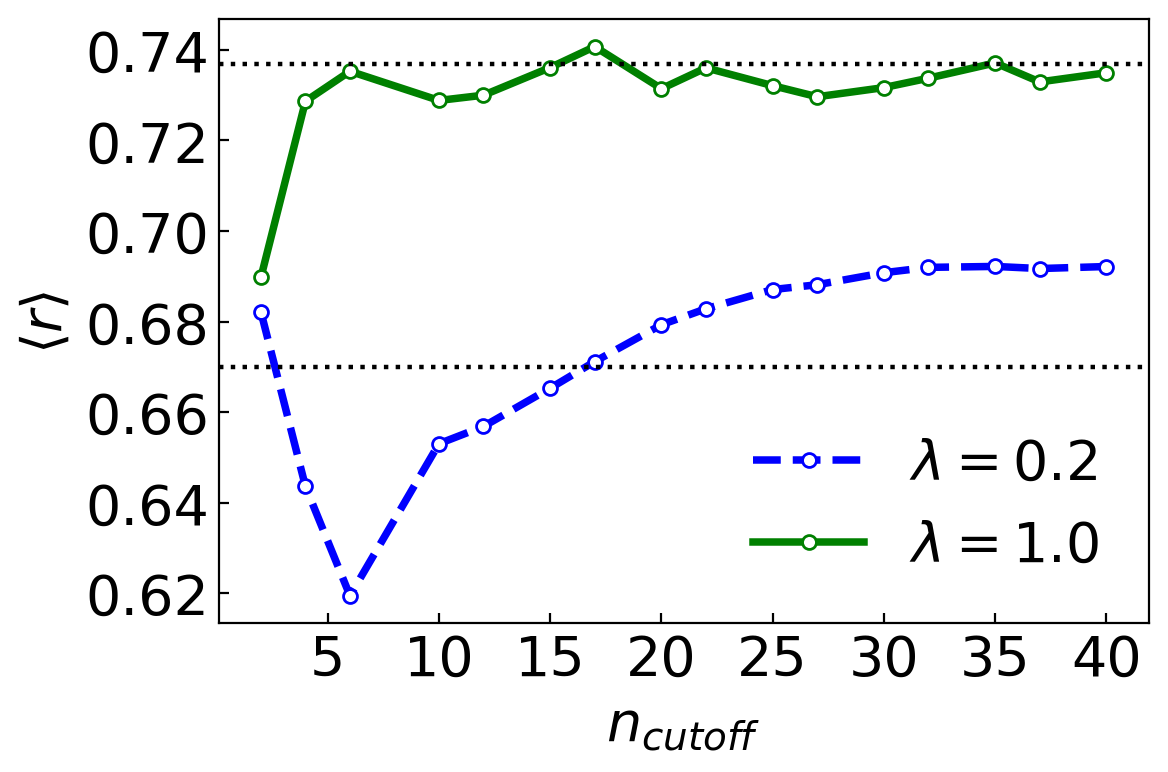}
\caption{$\langle \cos\theta\rangle$ and $\langle r \rangle$ as a function of $n_{\rm cutoff}$, computed from the consecutive complex level-spacing ratio distribution of $z$ introduced in Eq.~(7) in the main text ($S=10$). }
\label{NRSW_origin}
\end{figure}

\section{Convergence of the statistical properties of the spectrum with respect to the cavity cutoff}
\label{app_sec_conv_spec}
In this Section we discuss the convergence of the statistical properties of the spectrum of the Liouvillian $\mathcal{L}$ of the dissipative Dicke model with respect to the cavity cutoff, $n_{\rm cutoff}$, that was introduced in Section~\ref{app_sec_algo}. 

While the introduction of such a finite cutoff is essential to the numerical diagonalization of the Liouvillian, the repercussions on the resulting spectrum must be dealt with care.
In Fig.~\ref{Ncut_spectrum}, we plot the spectrum of $\mathcal{L}$ both in the normal and in the superradiant phase for different values of $n_{\rm cutoff}=20,30,40$.  
Note that we are only focusing on the window $\textrm{Re} \,E \in [ -\frac{2}{3}\times40\,\kappa, 0 ]$. 
In the normal phase, the three cutoffs yield the same highly patterned spectrum in the window $\textrm{Re} \,E \in [ -10 \,\kappa, 0 ]$. 
The patterned region of the spectrum grows as  $n_{\rm cutoff}$ is increased. 
For $n_{\rm cutoff}=40$, the whole window $\textrm{Re} \,E \in [ -\frac{2}{3}\times40\,\kappa, 0 ]$ is patterned. In the superradiant phase, convergence is obtained in the window $\textrm{Re} \,E \in [ -5 \,\kappa, 0 ]$. 

Rather than convergence of the location of eigenvalues in the complex plane, it is more important to ensure the convergence of spectral statistics.  In Fig.~\ref{NRSW_origin}, we follow the convergence of properties extracted from the consecutive level-spacing ratio distribution~\cite{sa2020complex} introduced in Eq.~(7) in the main text. Both $\langle r \rangle$ and $\langle \cos\theta \rangle$ remarkably converge when 
$n_{\rm cutoff} \approx 30$. All the results presented in the main text are produced with $n_{\rm cutoff} =40$.

\bibliography{references.bib}